\documentclass[11pt,a4paper]{article}
\usepackage[T1]{fontenc}\usepackage[utf8]{inputenc}
\usepackage[margin=2.7cm]{geometry}
\usepackage{amsmath,amssymb,amsthm,bm,mathtools,graphicx,xcolor}
\usepackage[colorlinks,linkcolor=blue!55!black,citecolor=blue!55!black,urlcolor=blue!55!black]{hyperref}
\usepackage{tikz-cd}
\usepackage{placeins}
\newcommand{\CC}{\mathbb{C}}
\newcommand{\RR}{\mathbb{R}}
\newcommand{\CP}{\mathbb{CP}}
\newcommand{\ZZ}{\mathbb{Z}}
\newcommand{\ii}{\mathrm{i}}
\newcommand{\dd}{\mathrm{d}}
\newcommand{\Id}{\mathbf{1}}
\newcommand{\Eperp}{E_{\!\perp}}
\newcommand{\Epar}{E_{\!\parallel}}
\newcommand{\dbar}[1]{\bar{\bar{#1}}}
\newcommand{\SFF}{\mathsf{B}}
\DeclareMathOperator{\im}{im}\DeclareMathOperator{\Hom}{Hom}
\DeclareMathOperator{\End}{End}
\theoremstyle{plain}
\newtheorem{theorem}{Theorem}\newtheorem{proposition}[theorem]{Proposition}
\newtheorem{lemma}[theorem]{Lemma}\newtheorem{corollary}[theorem]{Corollary}
\theoremstyle{definition}
\newtheorem{definition}[theorem]{Definition}\newtheorem{remark}[theorem]{Remark}

\begin{document}

\title{\bfseries The polarization bundle of a bending ray:\\
nested projections and helicity Chern numbers}
\author{Didier Felbacq\thanks{Laboratoire Charles Coulomb, Université de Montpellier, CNRS,
F-34095 Montpellier, France.} \and Emmanuel Rousseau\footnotemark[1]}
\date{\today}\maketitle

\begin{abstract}
We consider a ray propagating in a weakly anisotropic, possibly absorbing medium. The transverse
planes of the propagation directions form a rank-two complex vector bundle $\Eperp$ over the sphere
$S^2$ of directions, and the polarization states are the points of its projectivisation $Z=\CP(\Eperp)$.
Projecting the flat derivative of $\CC^3$ onto $\Eperp$ gives a connection $D$, the complexified
Levi-Civita connection of $S^2$, whose holonomy is the Rytov rotation. Projecting $D$ onto the line
of a polarization state gives the connection of the tautological line bundle over $Z$, whose
holonomy is the Pancharatnam--Berry phase. With respect to the horizontal distribution defined by
$D$, the curvature of this line bundle is $\ii\omega_{\mathrm{FS}}$ on the vertical part ($\omega_{\mathrm{FS}}$ the
Fubini--Study form), $\ii s_3\,\omega$ on the horizontal part ($\omega$ the area form of $S^2$, $s_3$ the circular
Stokes parameter), and zero on the mixed part. The Rytov rotation, the Pancharatnam--Berry phase and the Chern numbers
$c_1(L_\pm)=\mp2$ are thus read from one curvature. Restricted to the helicity sections, it gives the
obstruction to a globally smooth circular Jones vector, although the circular polarization states
are globally defined.
\end{abstract}
\section{Introduction and setting}

We investigate the propagation of light in a weakly anisotropic medium. 
Let $L$ be a characteristic length over which the background medium varies, and set
\begin{equation}
    \eta=\frac{1}{k_0L}\ll 1.
    \label{eq:eta}
\end{equation}
We introduce the dimensionless variable
\[
    \bm x=\frac{\bm r}{L}.
\]
The dielectric tensor is assumed to have the form
\begin{equation}
    \varepsilon(\bm x)
    =n_0^2(\bm x)\Id+\eta\,\chi(\bm x),
    \label{eq:epsilon}
\end{equation}
%
where $n_0$ is the index of the isotropic background, $k_0$ is the free-space wavenumber and $\chi$
is a bounded matrix-valued function, not assumed Hermitian: its Hermitian part is the birefringence
of the medium and its anti-Hermitian part the dichroism. The factor $\eta$ characterizes the
weakness of the anisotropic medium: the anisotropic and absorptive parts of the dielectric tensor are small
enough that they do not split the ray, so that the two polarizations follow one and the same
trajectory, determined by the $n_0$ alone, and act only as endomorphisms of the
transverse plane carried along it. This excludes strong double refraction, where the two
eigen-polarizations follow distinct rays and the very notion of a single transverse plane fails.
Here, $\chi=O(1)$ as $\eta\to0$. Thus birefringence and dichroism enter at
the transport order of geometrical optics.

We write $\hat{\bm k}(s)$ for the unit tangent of such a ray at arc length $s$. At each point
of the ray the amplitude of the field belongs to the complex plane orthogonal to
$\hat{\bm k}(s)$. As the ray bends, that plane moves. To study this motion, we do not follow one
ray but consider at once the transverse planes of all possible directions: they form a rank-two
complex vector bundle over the sphere of directions. A particular ray then selects a path
$s\mapsto\hat{\bm k}(s)$ in that sphere, along which its polarization is transported. The base
space is therefore the sphere $S^{2}$, although light propagates in physical space.

The geometry of polarization transport is usually presented either in coordinates, where it reduces
to a connection one-form on the sphere of ray directions~\cite{Rytov1938,Vladimirskii1941}, or in the
language of Berry's phase~\cite{Berry1984,Simon1983}, where it appears as the holonomy of a line
bundle. However, neither formulation makes clear that two distinct projections are in fact involved.
The first one comes from the transversality condition imposed on the field. It selects a rank-two
subbundle of the bundle of field amplitudes. The second projection is the passage to polarization
states: a polarization state is in fact a line, and this amounts to projectivizing that subbundle.
We consider both projections together and then compute what each of them discards.
The result is a curvature splitting on the full
space of polarization states $Z=\CP(\Eperp)$. Relative to the horizontal distribution
induced by transversality, the curvature of the tautological line bundle has vertical
part $F_{VV}=\ii\omega_{\mathrm{FS}}$, horizontal part $F_{HH}=\ii s_3\omega$,
and vanishing mixed part $F_{HV}=0$. The Pancharatnam and Rytov contributions thus arise
as two parts of one curvature. This is Theorem~\ref{thm:split}. The vanishing of the mixed part
does not depend on a choice of coordinates: it holds because the horizontal distribution is
defined by $D$, so that the projector onto a state has zero covariant derivative along horizontal
vectors. This mechanism is robust: it holds for the tautological connection induced by any Hermitian connection, and here
the connection is fixed by transversality. The helicity enters in the value of $F_{HH}$.

The joint evolution of the propagation direction and of the polarization state in a weakly
anisotropic inhomogeneous medium has been studied before. Liberman and
Zel'dovich~\cite{LibermanZeldovich1992} obtained the spin--orbit coupling of a photon in an
inhomogeneous medium; Bliokh, Frolov and Kravtsov~\cite{BliokhFrolovKravtsov2007} derived the
non-Abelian evolution of the Stokes vector in a lossless weakly anisotropic medium, with both the bending of the
ray and the anisotropy present, in terms of the Berry connection and curvature on the sphere of directions; the two effects were brought together and observed by
Bliokh, Niv, Kleiner and Hasman~\cite{Bliokh2008}, and the field is reviewed
in~\cite{BliokhSOI2015}. The asymptotic theory of geometric phases for coupled wave equations, of
which the present reduction is an instance, is due to Littlejohn and
Flynn~\cite{LittlejohnFlynn1991}. Bliokh unifies the Pancharatnam--Berry and the spin-redirection phases in the dynamical form
$\Phi_{\rm G}=-\int\bm S\cdot\bm\Omega_\zeta\,\dd\zeta$, with $\bm S$ the spin angular momentum and $\bm\Omega_\zeta$ the angular velocity of the
frame~\cite{Bliokh2009,BliokhSOI2015}, and describes the Berry curvature on the sphere of directions as the curvature of a surface of
unit Gaussian curvature~\cite{Bliokh2009}. Here the two phases are the vertical and the horizontal parts of the curvature of one
connection on the space of polarization states. The geometric status of this evolution is the subject of the present work.

\paragraph{Position of the result.} We build on the transport law of Rytov and Vladimirskii in its Stokes-vector form,
on the sub-bundle geometry of Oancea, Mieling and Palumbo~\cite{OanceaMielingPalumbo2026} (projected connections and
generalized Gauss--Codazzi--Mainardi equations, of which Lemma~\ref{lem:proj} and Corollary~\ref{cor:flatcase} are the form used
here), and on the minimal-coupling structure of Sternberg~\cite{Sternberg1977} and Weinstein~\cite{Weinstein1978}
(see also Montgomery~\cite{Montgomery1988} for its use in adiabatic phases). Applying this geometry to the two nested
sub-bundles defined by Maxwell's equations, $\Eperp\subset S^2\times\CC^3$ and $\mathcal L\subset\pi^{*}\Eperp$ over $Z$,
we obtain the following.
\begin{enumerate}
\item The connection $D$ is the complexified Levi-Civita connection of $S^2$, obtained from transversality by the Gauss
equation (Proposition~\ref{prop:curv}), and its holonomy is the Rytov rotation (Corollary~\ref{cor:hol}).
\item The curvature of the tautological line bundle is defined and given in closed form over the whole space $Z$ of polarization
states, that is, the direction and the Poincar\'e sphere together, with its vertical, horizontal and mixed parts relative to the
horizontal distribution of bending (Theorem~\ref{thm:split}). It does not depend on $\chi$, so that it holds for Hermitian
and non-Hermitian $\chi$. The moment map in the horizontal part is the circular Stokes parameter.
\item The Rytov rotation, the Pancharatnam phase, the Chiao--Wu phase and the two helicity Chern numbers are read from this
single curvature (\S5.6 and Section~\ref{sec:global}).
\item As a consequence of the vanishing of the mixed curvature, the closed circuit that bends, acts, unbends and
undoes has a geometric phase $O(\varepsilon\delta^2)$ (Proposition~\ref{prop:commute}). The circular polarization states are globally defined
while the corresponding Jones vectors are not (Corollary~\ref{cor:global}).
\end{enumerate}

\begin{figure}[t]\centering
\includegraphics[width=0.94\textwidth]{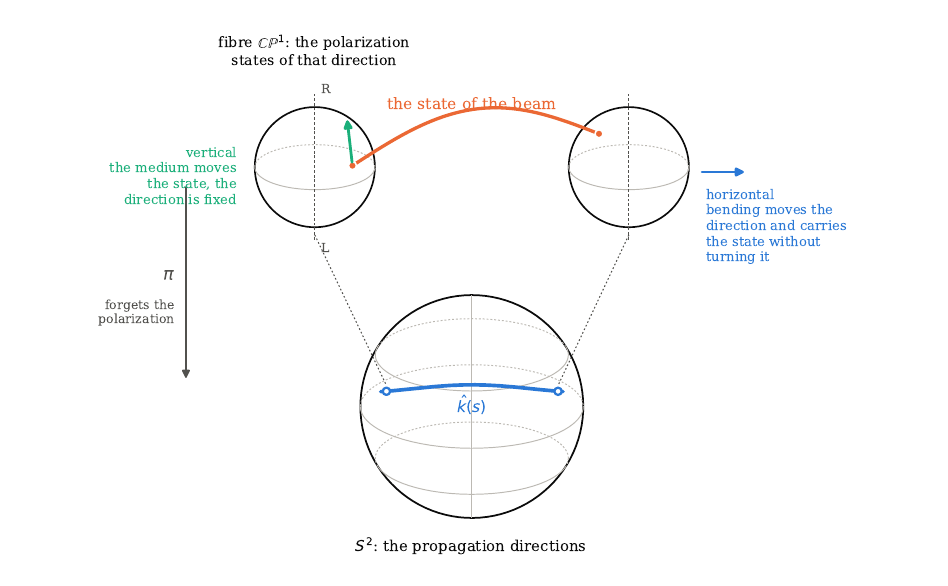}
\caption{The object of this work. A point of the total space is a propagation direction together
with a polarization state \emph{of that direction}; the beam is a curve of such points. Moving
inside a fibre is what the medium does, at fixed direction. Moving from one fibre to another, while
carrying the state without turning it in its plane, is what bending does --- but saying what
``without turning it'' means requires a rule, which Section~\ref{sec:transv} supplies. The curvature
computed in Section~\ref{sec:curv} measures the phase acquired around a small closed circuit of beam
states. In particular, the mixed curvature vanishes: a small circuit made of one bending step of size $\varepsilon$ and
one medium step of size $\delta$, followed by their inverses and closed in the fibre, acquires a phase
$O(\varepsilon\delta^2)$ (Proposition~\ref{prop:commute}).}
\label{fig:bundle}
\end{figure}

The helicity Chern numbers $c_1(L_\pm)=\mp2$ were obtained by Palmerduca and
Qin~\cite{PalmerducaQin2024,PalmerducaQin2025} for the photon bundle over the forward light cone
in free space, from the Berry curvature $\pm\ii\,\omega$ of the helicity lines. The fibre of their bundle at $\bm k$ is the
transverse plane $\bm k^\perp\otimes\CC$, so that, after radial projection of the cone onto $S^2$, it is the bundle $\Eperp$
considered here; equivalently $\Eperp=TS^2\otimes\CC$ (Remark~\ref{rem:TS2}). In Section~\ref{sec:global} these two curvatures
appear as the restrictions to the two helicity sections of the single curvature of Theorem~\ref{thm:split}, which also yields
the Rytov and Pancharatnam phases. The obstruction to a smooth global circular Jones frame is the Poincar\'e--Hopf
obstruction, obtained in Corollary~\ref{cor:global}. The geometrical construction used throughout is depicted in
Figure~\ref{fig:bundle}.

\paragraph{Geometric setting and notation.}
We denote
\begin{equation}
 M=S^{2}=\{\hat k\in\RR^{3}:|\hat k|=1\},\qquad E=M\times\CC^{3},
 \label{eq:ambient}
\end{equation}
with the standard Hermitian inner product $\langle\cdot,\cdot\rangle$ on $\CC^{3}$, conjugate-linear
in the first slot. The constant Cartesian frame gives a componentwise derivative on sections of
$E$; from Section~\ref{sec:transv} onward we denote it by $\nabla=\dd$. In
Section~\ref{sec:transport}, by contrast, $\nabla$ denotes differentiation with respect to the
dimensionless coordinate $\bm x$.

For later use we write $\Gamma(F)$ for the smooth sections of a complex vector bundle $F\to M$ and
$\Omega^{k}(M;F)$ for the $F$-valued $k$-forms on $M$. Throughout, every operator is given with its
source and target, because the objects at hand --- vector-valued functions, endomorphism-valued
forms, maps between different bundles --- are easy to confuse. We write
$X\cdot\Phi=\sum_a X^a\Phi^a$ for the bilinear (not sesquilinear) pairing of $X\in\RR^{3}$ with
$\Phi\in\CC^{3}$, and $\omega$ for the area form of the unit sphere, oriented by the outward normal
$\hat k$, so that $\int_{S^2}\omega=4\pi$. Spherical coordinates $(\theta,\varphi)$ on the direction
sphere carry the frame $\hat e_\theta=\partial_\theta\hat k$,
$\hat e_\varphi=(\sin\theta)^{-1}\partial_\varphi\hat k$, for which
$(\hat e_\theta,\hat e_\varphi,\hat k)$ is right-handed.
\FloatBarrier

\section{Geometrical-optics reduction}\label{sec:transport}

We now derive from Maxwell's equations the three facts used in the geometrical construction below:
the transverse plane, the ray flow, and the transport law inside that moving plane.
The derivation is a formal high-frequency expansion on a smooth branch of the eikonal $S$, away from
caustics. It is carried out at the level of the principal symbol and of the solvability condition at the next order.
The expansion belongs to the general asymptotic theory of coupled wave equations~\cite{LittlejohnFlynn1991}; for Maxwell's
equations in a smoothly inhomogeneous anisotropic medium the corresponding evolution of the polarization is
derived in~\cite{BliokhFrolovKravtsov2007}. In the guided case of curved step-index optical fibres,
Mieling and Oancea~\cite{MielingOancea2023} derive Rytov's law at leading order and the next correction.

We consider the high-frequency ansatz
\begin{equation}
 \bm E(\bm x)=e^{\ii S(\bm x)/\eta}\bm A(\bm x),\qquad
 \bm A=\bm a_0+\frac{\eta}{\ii}\,\bm a_1+\cdots,
 \label{eq:wkb}
\end{equation}
with $\eta\to0$, and we denote $\bm\xi=\nabla S$, where in this section $\nabla=\nabla_{\bm x}$ is the gradient with respect to the
dimensionless variable $\bm x$. We return to the physical variable $\bm r=L\bm x$ in \eqref{eq:fermat}.

The scaling chosen in \eqref{eq:epsilon} is important. The anisotropic term is one order below the
isotropic one: it does not change the principal symbol, the characteristic set or the rays, but it
does enter the transport equation. This is exactly what is needed here. Bending determines how the
transverse plane itself is carried, while $\chi$ acts inside that plane.

The expansion gives the following three statements.
\begin{proposition}\label{prop:main}
\begin{enumerate}
\item At each point of a ray the amplitude of the field is constrained to the complex plane
orthogonal to the direction of propagation. That is,
\begin{equation}
 \bm\xi\cdot\bm a_0=0,\qquad |\bm\xi|^{2}=n_0^{2}.
 \label{eq:eikonal}
\end{equation}
In the following, we denote by $P$ the orthogonal projector onto that plane.

\item The rays are the trajectories of a Hamiltonian flow in phase space, and satisfy Fermat's
equation \eqref{eq:fermat}.

\item Let $A_{\rm g}(\sigma)>0$ satisfy
\begin{equation}
 2n_0\,\frac{\dd A_{\rm g}}{\dd\sigma}+(\Delta S)A_{\rm g}=0,
 \label{eq:scalaramp}
\end{equation}
and write $\bm a_0=A_{\rm g}\bm e$. Then $\bm e$ is transverse and satisfies
\begin{equation}
 P\,\frac{\dd\bm e}{\dd\sigma}=\frac{\ii}{2n_0}\,P\chi P\,\bm e ,
 \label{eq:fulltransport}
\end{equation}
where $\sigma$ is arc length in the dimensionless variable $\bm x$.
The equation \eqref{eq:fulltransport} separates two effects. The term
\[
P\frac{\dd\bm e}{\dd\sigma}
\]
comes from the motion of the transverse plane along the bent ray, i.e. the geometrical transport. The term
\[
P\chi P\,\bm e
\]
describes the action of the anisotropic and absorptive part of the medium
within that plane.
\end{enumerate}
\end{proposition}
\begin{proof}
After multiplication by $L^2$, the time-harmonic Maxwell equation reads
\begin{equation}
    \nabla\times\nabla\times\bm E
    =
    \frac{1}{\eta^2}n_0^2\bm E
    +\frac{1}{\eta}\chi\bm E.
    \label{eq:maxwell}
\end{equation}
The identity
\[
\nabla\times(f\bm A)
=
f\,\nabla\times\bm A+\nabla f\times\bm A
\]
applied to $f=e^{\ii S/\eta}$ gives
\begin{equation}
    \nabla\times\bm E
    =
    e^{\ii S/\eta}
    \left(
        \nabla\times\bm A
        +\frac{\ii}{\eta}\bm\xi\times\bm A
    \right),
    \label{eq:curl1}
\end{equation}
and a second application gives
\begin{equation}
    \nabla\times\nabla\times\bm E
    =
    e^{\ii S/\eta}
    \left(
        \nabla\times\nabla\times\bm A
        +\frac{\ii}{\eta}\mathcal L\bm A
        -\frac{1}{\eta^2}
        \bm\xi\times(\bm\xi\times\bm A)
    \right),
    \label{eq:curl2}
\end{equation}
where
\begin{equation}
    \mathcal L\bm a
    =
    \nabla\times(\bm\xi\times\bm a)
    +\bm\xi\times(\nabla\times\bm a).
\end{equation}
Using
\[
    \bm\xi\times(\bm\xi\times\bm A)
    =
    \bm\xi(\bm\xi\cdot\bm A)-|\bm\xi|^2\bm A,
\]
equation \eqref{eq:maxwell} becomes
\begin{equation}
    \frac{1}{\eta^2}\dbar{\sigma}\bm A
    +\frac{\ii}{\eta}\mathcal L\bm A
    -\frac{1}{\eta}\chi\bm A
    +\nabla\times\nabla\times\bm A
    =0,
    \label{eq:hierarchy}
\end{equation}
with
\begin{equation}
    \dbar{\sigma}(\bm x,\bm\xi)
    =
    \bigl(|\bm\xi|^2-n_0(\bm x)^2\bigr)\Id
    -\bm\xi\bm\xi^{\!\top}
    \in\End(\CC^3).
    \label{eq:sigmabar}
\end{equation}
The four terms in \eqref{eq:hierarchy} are respectively of order
$\eta^{-2}$, $\eta^{-1}$, $\eta^{-1}$ and $1$. We now take the first two
orders.

\paragraph{Order $\eta^{-2}$: the transverse plane.}
At leading order, \eqref{eq:hierarchy} gives
\begin{equation}
    \dbar{\sigma}\bm a_0=0.
    \label{eq:symbol}
\end{equation}
The matrix $\dbar{\sigma}$ is the principal symbol. The term $\chi$ is absent
at this order because of the scaling in \eqref{eq:epsilon}.

Taking the scalar product of \eqref{eq:symbol} with $\bm\xi$ gives
\[
    -n_0^2(\bm\xi\cdot\bm a_0)=0,
\]
hence
\[
    \bm\xi\cdot\bm a_0=0.
\]
Equation \eqref{eq:symbol} then reduces to
\[
    (|\bm\xi|^2-n_0^2)\bm a_0=0,
\]
so a non-zero leading amplitude also requires
\[
    |\bm\xi|^2=n_0^2.
\]
These are exactly the two conditions \eqref{eq:eikonal}. The set on which
$\dbar{\sigma}$ has a non-trivial kernel,
\begin{equation}
    \mathrm{Char}
    =
    \bigl\{(\bm x,\bm\xi):|\bm\xi|=n_0(\bm x)\bigr\},
    \label{eq:char}
\end{equation}
is the characteristic set, and at each of its points
\begin{equation}
    \ker\dbar{\sigma}(\bm x,\bm\xi)
    =
    \{\bm a\in\CC^3:\bm\xi\cdot\bm a=0\},
    \qquad
    \hat{\bm k}=\frac{\bm\xi}{n_0}.
    \label{eq:kersigma}
\end{equation}
This is the complexified plane orthogonal to $\hat{\bm k}$, of complex
dimension two.

\paragraph{The rays.}
The characteristic set \eqref{eq:char} is the zero level of
\[
    p(\bm x,\bm\xi)
    =
    \frac12\bigl(|\bm\xi|^2-n_0^2\bigr),
\]
whose Hamiltonian flow is
\begin{equation}
    \dot{\bm x}=\bm\xi,
    \qquad
    \dot{\bm\xi}=n_0\nabla n_0.
    \label{eq:hamflow}
\end{equation}
Let $\sigma$ denote arc length in the dimensionless variable $\bm x$.
On $p=0$, $|\bm\xi|=n_0$, hence
\[
    \dd\sigma=n_0\,\dd t,
    \qquad
    \frac{\dd\bm x}{\dd\sigma}
    =
    \frac{\bm\xi}{n_0}
    =
    \hat{\bm k}.
\]
Equation \eqref{eq:hamflow} then gives
\begin{equation}
    \frac{\dd}{\dd\sigma}
    \left(
        n_0(\bm x(\sigma))
        \frac{\dd\bm x}{\dd\sigma}
    \right)
    =
    \nabla n_0(\bm x(\sigma)).
    \label{eq:fermat_dimensionless}
\end{equation}
Returning to the physical variables
\[
    \bm r=L\bm x,
    \qquad
    s=L\sigma,
\]
this is
\begin{equation}
    \frac{\dd}{\dd s}
    \left(
        n_0(\bm r(s))
        \frac{\dd\bm r}{\dd s}
    \right)
    =
    \nabla_{\bm r}n_0(\bm r(s)),
    \qquad
    \hat{\bm k}(s)=\frac{\dd\bm r}{\dd s}.
    \label{eq:fermat}
\end{equation}
Since $\chi$ does not enter the principal symbol, it does not enter the ray
equation at this order.

\paragraph{Order $\eta^{-1}$: the transport equation.}
At the next order, $\chi$ does enter. Since
\[
    \bm A
    =
    \bm a_0+\frac{\eta}{\ii}\bm a_1+O(\eta^2),
\]
the terms of order $\eta^{-1}$ in \eqref{eq:hierarchy} give
\begin{equation}
    -\ii\dbar{\sigma}\bm a_1
    +\ii\mathcal L\bm a_0
    -\chi\bm a_0
    =0,
\end{equation}
or equivalently
\begin{equation}
    \dbar{\sigma}\bm a_1
    =
    \mathcal L\bm a_0+\ii\chi\bm a_0.
    \label{eq:order1}
\end{equation}
The matrix $\dbar{\sigma}$ is real and symmetric, hence Hermitian, so its
image is the orthogonal complement of its kernel. Equation
\eqref{eq:order1} has a solution $\bm a_1$ if and only if its right-hand
side is orthogonal to $\ker\dbar{\sigma}$. The solvability condition is
therefore
\begin{equation}
    P\bigl(\mathcal L\bm a_0+\ii\chi\bm a_0\bigr)=0,
    \qquad
    P=\Id-\hat{\bm k}\hat{\bm k}^{\!\top}.
    \label{eq:solv}
\end{equation}
Since $P\bm a_0=\bm a_0$, the material term can be written
$P\chi P\bm a_0$. Only the transverse block of $\chi$ enters the leading
transport equation.

The differential part is made explicit by
\begin{align}
    \nabla\times(\bm\xi\times\bm a)
    &=
    \bm\xi(\nabla\cdot\bm a)
    -(\Delta S)\bm a
    +(\bm a\cdot\nabla)\bm\xi
    -(\bm\xi\cdot\nabla)\bm a,\\
    \bm\xi\times(\nabla\times\bm a)
    &=
    \nabla(\bm\xi\cdot\bm a)
    -(\bm\xi\cdot\nabla)\bm a
    -(\bm a\cdot\nabla)\bm\xi,
\end{align}
where the second identity uses
$\nabla\times\bm\xi=\nabla\times\nabla S=0$. Hence
\begin{equation}
    \mathcal L\bm a
    =
    \bm\xi(\nabla\cdot\bm a)
    -(\Delta S)\bm a
    -2(\bm\xi\cdot\nabla)\bm a
    +\nabla(\bm\xi\cdot\bm a).
    \label{eq:Lexplicit}
\end{equation}
Take $\bm a=\bm a_0$ and apply $P$. The first term vanishes because
$P\bm\xi=0$, the last one because
$\bm\xi\cdot\bm a_0=0$, and $P\bm a_0=\bm a_0$. Since
\[
    \bm\xi=n_0\hat{\bm k},
    \qquad
    \hat{\bm k}\cdot\nabla
    =
    \frac{\dd}{\dd\sigma},
\]
equation \eqref{eq:solv} becomes
\begin{equation}
    2n_0P\frac{\dd\bm a_0}{\dd\sigma}
    +(\Delta S)\bm a_0
    =
    \ii P\chi P\bm a_0.
    \label{eq:transport_dimensionless}
\end{equation}

The scalar spreading of the ray can now be separated from the polarization.
Let $A_{\rm g}(\sigma)>0$ solve
\begin{equation}
    2n_0\frac{\dd A_{\rm g}}{\dd\sigma}
    +(\Delta S)A_{\rm g}=0
\end{equation}
and write
\[
    \bm a_0=A_{\rm g}\bm e.
\]
When $\chi$ has an anti-Hermitian part, $\bm e$ need not have constant norm;
its complex line is the polarization state. Substitution into
\eqref{eq:transport_dimensionless} gives
\begin{equation}
    P\frac{\dd\bm e}{\dd\sigma}
    =
    \frac{\ii}{2n_0}P\chi P\bm e ,
\end{equation}
In terms of the physical arc length $s=L\sigma$, this becomes
\begin{equation}
    P\frac{\dd\bm e}{\dd s}
    =
    \frac{\ii}{2n_0L}P\chi P\bm e.
    \label{eq:fulltransport-s}
\end{equation}

If $\chi$ is Hermitian, the action of $P\chi P$ is norm-preserving
and describes birefringence. Its anti-Hermitian part describes differential
attenuation or amplification.

When the material term is absent, \eqref{eq:fulltransport} reduces to
\begin{equation}
    P\frac{\dd\bm e}{\dd s}=0.
    \label{eq:rytovtransport}
\end{equation}
This is the transport used below to define the horizontal motion. The
material term gives the motion inside the polarization fibre.
\end{proof}

\section{The space of polarization states}\label{sec:space}

A beam propagating in a weakly anisotropic medium carries both its
direction of propagation and its state of polarization. These quantities are not independent and evolve together. The polarization lives in the plane transverse to the direction, so when the direction turns the plane turns with it.
The medium acts inside that plane and moves the polarization without touching the direction. The
object that describes both quantities at once is therefore a pair
\begin{equation}
 \big(\hat k,\ \ell\big),\qquad
 \hat k\in S^{2},\quad \ell\ \text{a polarization state of the plane}\ \Eperp(\hat k) ,
 \label{eq:state}
\end{equation}
and the propagation of the beam induces a curve of such pairs. 

Explicitly, let $\hat{\bm k}(s)$ be the unit tangent of the ray, given by \eqref{eq:fermat}, and let
\begin{equation}
 U(s):\ \Eperp\big(\hat{\bm k}(0)\big)\longrightarrow\Eperp\big(\hat{\bm k}(s)\big)
 \label{eq:propagator}
\end{equation}
be the linear map that takes $\bm a_0(0)$ to $\bm a_0(s)$, obtained by solving the transport
equation \eqref{eq:transport_dimensionless} along the ray. Equation \eqref{eq:transport_dimensionless} is linear in $\bm a_0$,
so $U(s)$ maps complex lines to complex lines, and the curve is
\begin{equation}
 s\longmapsto\gamma(s)=\big(\hat{\bm k}(s),\ \ell(s)\big),
 \qquad \ell(s)=U(s)\,\ell(0) .
 \label{eq:curveZ}
\end{equation}
The first component is fixed by the ray alone and the second by $\ell(0)$ together with the medium,
and $\pi\circ\gamma=\hat{\bm k}$. Neither the amplitude nor the phase of a vector spanning
$\ell(0)$ enters \eqref{eq:curveZ}, since $U(s)$ is linear: the curve in $Z$ depends on the initial
polarization state and not on the initial Jones vector.

This section builds the space of those
pairs and says, in words, what can move in it. Nothing is computed here.

\paragraph{Transverse planes and polarization states.} For each direction $\hat k$ let
\begin{equation}
 \Eperp(\hat k)=\hat k^{\perp}\otimes\CC
 \label{eq:Eperpdef}
\end{equation}
be the complexified plane orthogonal to it: by \eqref{eq:eikonal}, the amplitudes a field travelling
in that direction may have. As $\hat k$ runs over the sphere these planes form a rank-two complex
vector bundle $\Eperp\to S^{2}$, a sub-bundle of the trivial bundle $E=S^{2}\times\CC^{3}$ of all
amplitudes. Inside one plane, two amplitudes differing by a complex factor describe the same
polarization --- same ellipse, same handedness, a different intensity and a different origin of
phase --- so the \emph{states} available to that direction are the complex lines of
$\Eperp(\hat k)$, and they form a sphere: the Poincaré sphere of that direction.

\begin{definition}\label{def:Z}
Let
\begin{equation}
 Z=\CP(\Eperp)\ \xrightarrow{\ \pi\ }\ S^{2}
 \label{eq:Z}
\end{equation}
be the fibrewise projectivisation of the bundle of transverse planes: a point of $Z$ over $\hat k$
is a complex line $\ell\subset\Eperp(\hat k)$, that is, a polarization state of that direction, the
overall amplitude and phase discarded. The fibre $\CP\big(\Eperp(\hat k)\big)\cong\CP^{1}$ is the
Poincaré sphere of the direction $\hat k$, and $\pi:Z\to S^{2}$, $\pi(\ell)=\hat k$, is the map
that forgets the polarization and remembers the direction. Its differential at a point is
$\dd\pi_\ell:T_\ell Z\to T_{\hat k}S^{2}$, surjective with two-dimensional kernel.
\end{definition}

The propagation of a beam is thus a curve $s\mapsto(\hat k(s),\ell(s))$ in $Z$
(Figure~\ref{fig:bundle}), projecting onto the path of directions, which the trajectory fixes on its
own.

\paragraph{The two motions.} A curve in $Z$ can move for two reasons, and they are the two
physical mechanisms. It can move inside one fibre, the direction being held fixed: that is what the
medium does, birefringence and dichroism acting in the transverse plane without touching $\hat k$.
They act on the fibre by M\"obius transformations of the Poincaré sphere, which are rotations only
in the absence of loss. Or the direction can move and the polarization be carried along, untouched
by any medium: that is what bending does, and it still moves the point of $Z$, because the Poincaré
sphere the point sits on is itself being carried to another direction.

The second motion has no meaning yet. To say that a polarization has been carried from one direction
to a neighbouring one \emph{without being turned} presupposes a rule comparing the transverse planes
of two different directions, and nothing so far supplies one. This is where the first projection
enters: Section~\ref{sec:transv} shows that transversality supplies that rule. The rule is unique
once one requires, as the transport law of Proposition~\ref{prop:main} does, that the transported
field does not rotate about the ray (Proposition~\ref{prop:unique}).

\paragraph{The tautological line bundle.} One thing has to be restored before anything can be computed.
Passing to states has discarded the amplitude and, with it, the phase. The phase is what we want to
follow, so it has to be put back --- not as a number attached to each state, which would presuppose
an origin, but as the one-dimensional space of amplitudes available in that state. Over each point
$\ell$ of $Z$ sits a complex line, namely $\ell$ itself; a point of the resulting line bundle is a
Jones vector, and $Z$ is that bundle with the complex scale forgotten.

One bookkeeping step first. The bundle $\Eperp$ lives over the direction sphere, whereas
the states live over $Z$, so the two cannot be compared until $\Eperp$ has been transported to $Z$.
That is what the pullback along $\pi$ does:
\[
\begin{tikzcd}
\pi^{*}\Eperp \arrow[r,"\widetilde\pi"] \arrow[d,"p"'] & \Eperp \arrow[d,"q"]\\
Z \arrow[r,"\pi"'] & S^{2}
\end{tikzcd}
\]
A point of $\pi^{*}\Eperp$ is a pair $(\ell,\Phi)$ with $\ell\in Z$ and $\Phi\in\Eperp(\pi(\ell))$,
so that $(\pi^{*}\Eperp)_\ell=\Eperp(\hat k)$ for $\hat k=\pi(\ell)$. Nothing new appears in the
fibres; only the base changes, the same transverse plane being attached now to every state lying
above $\hat k$ --- which is what is needed, since a state is one line inside that plane and there is
one such line for each point of the fibre.

\begin{definition}\label{def:taut}
The \emph{tautological} line bundle over $Z$ is the rank-one sub-bundle
$\mathcal L\subset\pi^{*}\Eperp$ whose fibre over a state is that state:
\begin{equation}
 \mathcal L_\ell=\ell\ \subset\ \Eperp(\hat k)=\big(\pi^{*}\Eperp\big)_\ell ,
 \qquad \ell\in Z,\ \hat k=\pi(\ell).
 \label{eq:taut}
\end{equation}
\end{definition}

A section of $\mathcal L$ is a genuine Jones vector field, and comparing the phases it takes at two
different states is what a connection on $\mathcal L$ does. Section~\ref{sec:curv} builds the only
one available --- but it cannot be built before Section~\ref{sec:transv} has supplied $D$.

\section{The connection induced by transversality}\label{sec:transv}

Section~\ref{sec:space} left one thing missing: a rule for
carrying a polarization from one direction to a neighbouring one. This section shows that
transversality supplies it, up to the absence of rotation about the ray, which is a property of the
transport law and not of transversality. We write transversality as a field of projectors (\S4.1);
observe that differentiating a transverse field splits into two blocks, a connection $D$ and a
second fundamental form $\SFF$, which are not postulated but are the two halves of one
derivative (\S4.2); and prove one lemma on the curvature of a projected connection (\S4.3), which
will be used again in Section~\ref{sec:curv} for the second projection. On the sphere of directions
everything is then explicit: $D$ is the Levi-Civita connection of the round sphere and its holonomy
is the Rytov rotation (\S4.4--\S4.5). The medium plays no role here.

\subsection{The transversality projector}

\begin{definition}\label{def:P}
For each $\hat k\in S^{2}$ let
\begin{equation}
 P(\hat k)=\Id-\hat k\,\hat k^{\!\top}\ \in\ \End(\CC^{3})
 \label{eq:P}
\end{equation}
be the orthogonal projector of $\CC^{3}$ onto the complexified plane $\hat k^{\perp}\otimes\CC$:
the operation that discards whatever longitudinal component a field vector may carry, and keeps the
two transverse ones. Thus $P$ is a map $S^{2}\to\End(\CC^{3})$, and for each $\hat k$ the
endomorphism $P(\hat k):\CC^{3}\to\CC^{3}$ satisfies $P^{2}=P=P^{\dagger}$. The bundle introduced in
\eqref{eq:Eperpdef} is therefore its image, and its orthogonal complement is the longitudinal line:
\begin{equation}
 \Eperp=\im P,\qquad \Epar=\ker P=\CC\hat k ,
 \label{eq:split}
\end{equation}
so that $E=\Eperp\oplus\Epar$ as an orthogonal direct sum of complex vector bundles over $S^{2}$, of
ranks two and one.
\end{definition}

Two remarks fix the status of \eqref{eq:P}. First, $P$ is a smooth field of projectors because
$\hat k\mapsto\hat k\hat k^{\!\top}$ is smooth on the sphere; hence $\Eperp$ is a genuine smooth
sub-bundle and not merely a family of planes. Second, $P$ is the projector for the \emph{Hermitian}
product of $\CC^{3}$ although it is written with a transpose and not a conjugate transpose: this is
because $\hat k$ is real. The distinction matters later, since the pairing $X\cdot\Phi$ that appears
throughout is bilinear.

\subsection{The induced connection and the second fundamental form}

A section $\Phi\in\Gamma(\Eperp)$ is a field of amplitudes, one in each transverse plane. Being also
a section of $E$, it is a $\CC^{3}$-valued function on the sphere, and can simply be differentiated:
for $X\in T_{\hat k}S^{2}$, $\nabla_X\Phi=\dd\Phi(X)\in\CC^{3}$ is the derivative of its three
Cartesian components along $X$, taken in the sphere. So
\begin{equation}
 \nabla_X:\ \Gamma(E)\longrightarrow\Gamma(E),\qquad
 \nabla_XP=\dd P(X):\ \CC^{3}\longrightarrow\CC^{3} ,
 \label{eq:types}
\end{equation}
the second because $P:S^{2}\to\End(\CC^{3})$ is itself a matrix-valued function, so that
$\nabla P\in\Omega^{1}\big(S^{2};\End(\CC^{3})\big)$. The point is that $\nabla_X$ does not map
$\Gamma(\Eperp)$ into $\Gamma(\Eperp)$: the plane in which $\Phi$ lives has itself moved. The failure is measured
exactly. Differentiating the identity $P\Phi=\Phi$ gives
\begin{equation}
 (\nabla_XP)\Phi+P\,\nabla_X\Phi=\nabla_X\Phi ,
 \qquad\text{i.e.}\qquad
 (\Id-P)\,\nabla_X\Phi=(\nabla_XP)\,\Phi .
 \label{eq:leak}
\end{equation}
The right-hand side involves no derivative of $\Phi$. So the longitudinal part of $\nabla_X\Phi$ is
\emph{pointwise linear} in $\Phi$: it is not a differential operator but a tensor. This is the
observation the rest of the section rests on, and it motivates the two definitions below.

\begin{definition}\label{def:DII}
For $X\in T_{\hat k}S^{2}$ and $\Phi\in\Gamma(\Eperp)$, set
\begin{equation}
 D_X\Phi=P\,\nabla_X\Phi
 \qquad\text{and}\qquad
 \SFF(X,\Phi)=(\Id-P)\,\nabla_X\Phi=(\nabla_XP)\Phi ,
 \label{eq:DII}
\end{equation}
the two blocks of $\nabla_X\Phi$ along $E=\Eperp\oplus\Epar$. Their types are
\begin{equation}
 D:\ \Gamma(\Eperp)\to\Omega^{1}(S^{2};\Eperp),\qquad
 \SFF:\ TS^{2}\otimes\Eperp\to\Epar ,
 \label{eq:DIItypes}
\end{equation}
so that $D_X:\Gamma(\Eperp)\to\Gamma(\Eperp)$ is a differential operator while, for each $\hat k$
and $X$, $\SFF(X,\cdot):\Eperp(\hat k)\to\Epar(\hat k)$ is a linear map between two
\emph{different} fibres. We call $D$ the \emph{induced connection} on $\Eperp$ and $\SFF$
the \emph{second fundamental form} of $\Eperp$ in $E$. We also write
\begin{equation}
 \SFF^{*}(X)=P\,(\nabla_XP)\,(\Id-P)\ :\ \Epar(\hat k)\longrightarrow\Eperp(\hat k)
 \label{eq:Wein}
\end{equation}
for the adjoint, or Weingarten, map, which goes the other way.
\end{definition}

\noindent The names are those of classical submanifold theory: \eqref{eq:DII} is the
decomposition of $\nabla_X\Phi$ along $E=\Eperp\oplus\Epar$, the first block kept, the second
discarded. What \eqref{eq:leak} adds is that discarding it is legitimate --- $D$ is a connection,
because $\SFF$ is tensorial and therefore cannot spoil the Leibniz rule:
\begin{equation*}
 D_X(f\Phi)=P\nabla_X(f\Phi)=(Xf)\,P\Phi+f\,P\nabla_X\Phi=(Xf)\Phi+f\,D_X\Phi .
\end{equation*}
Likewise $\SFF(X,f\Phi)=f\,\SFF(X,\Phi)$ by \eqref{eq:leak}, and
$\SFF(fX,\Phi)=f\,\SFF(X,\Phi)$ trivially, so $\SFF$ is a bundle map
$TS^{2}\otimes\Eperp\to\Epar$ and not an operator. That $D$ is metric follows from
$P^{\dagger}=P$ and the flatness of $\nabla$ with respect to $\langle\cdot,\cdot\rangle$. These are
the standard facts about connections induced on sub-bundles; see Kobayashi and
Nomizu~\cite[Vol.~II, Ch.~VII, \S3]{KobayashiNomizu} for the submanifold case, of which this is the
linear-algebra skeleton.

\begin{proposition}\label{prop:unique}
Let $D'$ be a connection on $\Eperp$ such that, along every smooth curve $\hat k(s)$ in $S^2$, every
$D'$-parallel section $\bm e(s)$ of $\Eperp$ satisfies $P\dot{\bm e}=0$, equivalently $\dot{\bm e}\in\CC\,\hat k$: the
derivative of the transported field is along the ray, so that there is neither rotation about the ray nor
change of the field within the transverse plane. Then $D'=D$. Conversely, $D$ has this property, and this is the property expressed by
\eqref{eq:rytovtransport}.
\end{proposition}

\begin{proof}
Differentiating $\hat k\cdot\bm e=0$ gives $\hat k\cdot\dot{\bm e}=-\dot{\hat k}\cdot\bm e$. If
$\dot{\bm e}=\lambda\hat k$, then $\lambda=\hat k\cdot\dot{\bm e}$, so $\dot{\bm e}=-(\dot{\hat k}\cdot\bm e)\hat k$,
and $P\dot{\bm e}=0$, that is, $\bm e$ is $D$-parallel. Thus $D$ and $D'$ have the same parallel
transport along every curve, hence the same covariant derivative. Conversely, if $P\dot{\bm e}=0$ then
$\dot{\bm e}=(\Id-P)\dot{\bm e}\in\CC\hat k$.
\end{proof}

\noindent Transversality alone does not give this property. For any smooth real one-form $\tau$ on $S^2$,
$D'_X=D_X+\tau(X)\,\hat k\times$ is also a connection on $\Eperp$, whose parallel fields rotate about the ray at the
rate $\tau$. Such a term would describe optical activity. Its absence from the transport law
\eqref{eq:rytovtransport} comes from the solvability condition \eqref{eq:solv}, that is, from Maxwell's equations
and not from transversality. This is the input that selects $D$.

Physically, \eqref{eq:leak} says what it costs to turn a ray. The field itself never acquires a
longitudinal component: transversality holds at every point. It is the \emph{derivative} of the
field that acquires one, and $\SFF$ is that component. Section~4.4 computes it.

\subsection{Curvature of a projected connection}

All the curvature computations below come from one lemma. We state it for a general orthogonal
projector field, so that Section~\ref{sec:curv} can use it again without change.

\begin{lemma}[curvature of a projected connection]\label{lem:proj}
Let $\mathcal E\to M$ be a complex vector bundle with a connection
$\nabla:\Gamma(\mathcal E)\to\Omega^{1}(M;\mathcal E)$, let
$\Pi\in\Gamma\big(\End(\mathcal E)\big)$ be a smooth field of orthogonal projectors, let
$\mathcal F=\im\Pi\subset\mathcal E$, and let
$\nabla^{\Pi}=\Pi\circ\nabla:\Gamma(\mathcal F)\to\Omega^{1}(M;\mathcal F)$ be the induced
connection. Then, as elements of $\Omega^{2}\big(M;\End(\mathcal F)\big)$,
\begin{equation}
 F_{\nabla^{\Pi}}=\Pi\,F_{\nabla}\,\Pi+\Pi\,(\nabla\Pi\wedge\nabla\Pi)\,\Pi ,
 \label{eq:projcurv}
\end{equation}
where $\nabla\Pi\in\Omega^{1}\big(M;\End(\mathcal E)\big)$ is the covariant derivative of $\Pi$ as
an endomorphism-valued one-form, $F_\nabla\in\Omega^{2}\big(M;\End(\mathcal E)\big)$, and $\wedge$
combines the wedge of forms with the composition of endomorphisms: on two tangent vectors,
$(\nabla\Pi\wedge\nabla\Pi)(X,Y)=(\nabla_X\Pi)(\nabla_Y\Pi)-(\nabla_Y\Pi)(\nabla_X\Pi)$.
\end{lemma}

\begin{proof}
Differentiating $\Pi^{2}=\Pi$ gives $(\nabla\Pi)\Pi+\Pi(\nabla\Pi)=\nabla\Pi$; multiplying on both
sides by $\Pi$ yields $2\,\Pi(\nabla\Pi)\Pi=\Pi(\nabla\Pi)\Pi$, hence
\begin{equation}
 \Pi\,(\nabla\Pi)\,\Pi=0 .
 \label{eq:PdPP}
\end{equation}
Let $s$ be a section of $\mathcal F$, so $\Pi s=s$. Then $\nabla^{\Pi}s=\Pi\nabla s$ and
\begin{equation*}
 (\nabla^{\Pi})^{2}s=\Pi\,\nabla\big(\Pi\nabla s\big)
 =\Pi\,(\nabla\Pi)\wedge\nabla s+\Pi\,\Pi\,\nabla^{2}s
 =\Pi\,(\nabla\Pi)\wedge\nabla s+\Pi\,F_\nabla\,s .
\end{equation*}
In the first term write $\nabla s=\Pi\nabla s+(\Id-\Pi)\nabla s$. The first piece is killed by
\eqref{eq:PdPP}. For the second, $(\Id-\Pi)\nabla s=(\Id-\Pi)\nabla(\Pi s)=(\Id-\Pi)(\nabla\Pi)s$,
again by $\Pi s=s$ and $(\Id-\Pi)\Pi=0$. Hence, using \eqref{eq:PdPP} once more to replace
$\Pi(\nabla\Pi)(\Id-\Pi)$ by $\Pi(\nabla\Pi)$, and $s=\Pi s$ on the right,
\begin{equation*}
 \Pi\,(\nabla\Pi)\wedge\nabla s=\Pi\,(\nabla\Pi)\wedge(\nabla\Pi)\,\Pi\,s ,
\end{equation*}
which is \eqref{eq:projcurv}.
\end{proof}

\begin{corollary}\label{cor:flatcase}
If the ambient connection is flat, $F_\nabla=0$, then in terms of the second fundamental form and
its adjoint,
\begin{equation*}
 \SFF(X)=(\Id-\Pi)(\nabla_X\Pi):\ \mathcal F\to\mathcal F^{\perp},\qquad
 \SFF^{*}(X)=\Pi(\nabla_X\Pi)(\Id-\Pi):\ \mathcal F^{\perp}\to\mathcal F ,
\end{equation*}
one has, for $X,Y$ tangent to $M$ and as endomorphisms of $\mathcal F$,
\begin{equation}
 F_{\nabla^{\Pi}}(X,Y)=\SFF^{*}(X)\,\SFF(Y)-\SFF^{*}(Y)\,\SFF(X) .
 \label{eq:gauss}
\end{equation}
\end{corollary}

\begin{proof}
By \eqref{eq:PdPP}, $\Pi(\nabla_X\Pi)(\nabla_Y\Pi)\Pi
=\Pi(\nabla_X\Pi)(\Id-\Pi)(\nabla_Y\Pi)\Pi=\SFF^{*}(X)\SFF(Y)$; antisymmetrising in
$X,Y$ gives \eqref{eq:gauss} from \eqref{eq:projcurv}.
\end{proof}

\noindent Equation \eqref{eq:gauss} is the Gauss equation of classical submanifold theory
\cite[Vol.~II, Ch.~VII, \S4]{KobayashiNomizu}, here in its sub-bundle form and with the sign fixed
by the convention \eqref{eq:DII}. Equation \eqref{eq:projcurv} is the more general statement, needed
in Section~\ref{sec:curv} because the connection ambient to the \emph{second} projection is $D$,
which is not flat.
Stating the lemma once, in the general form, is what allows the two projections to be treated as one
construction.

\subsection{Curvature on the direction sphere}

Now specialise to \eqref{eq:P}. Regard $\hat k$ as the position field of $S^{2}\subset\RR^{3}$, so
that $\nabla_X\hat k=X$ for $X\in T_{\hat k}S^{2}$. Then \eqref{eq:P} gives at once
\begin{equation}
 \nabla_XP=-\big(X\,\hat k^{\!\top}+\hat k\,X^{\!\top}\big).
 \label{eq:dP}
\end{equation}

\begin{lemma}\label{lem:II}
For every $X\in T_{\hat k}S^{2}$, every $\Phi\in\Eperp(\hat k)$ and every $\zeta\in\CC$,
\begin{equation}
 \SFF(X,\Phi)=-(X\cdot\Phi)\,\hat k ,
 \qquad
 \SFF^{*}(X)\big(\zeta\hat k\big)=-\zeta\,X .
 \label{eq:IIexplicit}
\end{equation}
\end{lemma}

\begin{proof}
Apply $\Id-P=\hat k\hat k^{\!\top}$ on the left of \eqref{eq:dP} and $\Phi$ on the right. Using
$\hat k^{\!\top}X=0$, $\hat k^{\!\top}\hat k=1$ and $X^{\!\top}\Phi=X\cdot\Phi$, the term
$\hat k\hat k^{\!\top}X\hat k^{\!\top}\Phi$ vanishes and the term
$\hat k\hat k^{\!\top}\hat kX^{\!\top}\Phi$ gives $(X\cdot\Phi)\hat k$, with an overall minus sign.
For the second identity, $(\nabla_XP)(\zeta\hat k)=-\zeta\big(X(\hat k^{\!\top}\hat k)
+\hat k(X^{\!\top}\hat k)\big)=-\zeta X$, which is already transverse, so applying $P$ changes nothing.
\end{proof}

\begin{proposition}\label{prop:curv}
The curvature $F_D\in\Omega^{2}\big(S^{2};\End(\Eperp)\big)$ of the induced connection is given, for
$X,Y\in T_{\hat k}S^{2}$ and $\Phi\in\Eperp(\hat k)$, by
\begin{equation}
 F_D(X,Y)\,\Phi=(Y\cdot\Phi)\,X-(X\cdot\Phi)\,Y\ \in\ \Eperp(\hat k) .
 \label{eq:FD}
\end{equation}
Consequently $D$ is the complexification of the Levi-Civita connection of the round unit sphere, and
$F_D$ is its Riemann tensor.
\end{proposition}

\begin{proof}
The ambient $\nabla$ is flat, so Corollary~\ref{cor:flatcase} applies. By \eqref{eq:IIexplicit},
$\SFF^{*}(X)\SFF(Y)\Phi=\SFF^{*}(X)\big(-(Y\cdot\Phi)\hat k\big)
=(Y\cdot\Phi)X$, and antisymmetrising gives \eqref{eq:FD}. Restricted to real sections,
\eqref{eq:FD} is the curvature tensor $R(X,Y)Z=\langle Y,Z\rangle X-\langle X,Z\rangle Y$ of a space
form of sectional curvature $1$; and the connection induced on $TS^{2}$ inside
$TS^{2}\oplus\nu=\RR^{3}$ by the same construction is the Levi-Civita connection, by the Gauss
formula \cite[Vol.~II, Ch.~VII, \S3]{KobayashiNomizu}. The complexification of a real connection has
the complexified curvature, whence the statement.
\end{proof}

\subsection{Connection one-form and holonomy}

Proposition~\ref{prop:curv} is coordinate-free; to compare it with what is measured, write it in the
frame $(\hat e_\theta,\hat e_\varphi)$.

\begin{proposition}\label{prop:form}
Let $U\subset S^{2}$ be the chart of spherical coordinates, poles excluded, on which the frame
$(\hat e_\theta,\hat e_\varphi)$ is defined. In that frame $D$ reads
\begin{equation}
 D=\dd-A\,J ,\qquad
 A=-\cos\theta\,\dd\varphi\ \in\ \Omega^{1}(U;\RR) ,\qquad
 \dd A=\sin\theta\,\dd\theta\wedge\dd\varphi=\omega ,
 \label{eq:A}
\end{equation}
where $J=\left(\begin{smallmatrix}0&-1\\1&0\end{smallmatrix}\right)$ generates the rotations of the
plane, so that the connection form proper is the $\mathfrak{so}(2)$-valued one-form $-A\,J$ and $A$
is its single real component. Thus $A$ is the vector potential of a magnetic monopole of unit
charge, in the gauge regular away from the poles, the potential of the momentum-space monopole of
Refs.~\cite{BliokhFrolovKravtsov2007,Bliokh2008,Bliokh2009} written in the real frame $(\hat e_\theta,\hat e_\varphi)$.
\end{proposition}

\begin{proof}
From $\hat e_\theta=(\cos\theta\cos\varphi,\cos\theta\sin\varphi,-\sin\theta)$ and
$\hat e_\varphi=(-\sin\varphi,\cos\varphi,0)$ one computes
$\langle\dd\hat e_\theta,\hat e_\varphi\rangle=\cos\theta\,\dd\varphi$. Hence for a transverse unit
vector written as $\hat u(\psi)=\cos\psi\,\hat e_\theta+\sin\psi\,\hat e_\varphi$ the equation
$D\hat u=0$ reads
\begin{equation}
 \dd\psi=-\cos\theta\,\dd\varphi=A ,
 \label{eq:dpsi}
\end{equation}
which identifies $A$; and $\dd A=\sin\theta\,\dd\theta\wedge\dd\varphi$.
\end{proof}

\begin{corollary}[the Rytov rotation]\label{cor:hol}
Let $\gamma=\partial S$ be a closed loop on the direction sphere bounding a region $S$ of solid angle
$\Omega_{\mathrm{sol}}(\gamma)=\int_S\omega$, oriented compatibly. Parallel transport of a transverse
vector around $\gamma$ increases its angle in the frame $(\hat e_\theta,\hat e_\varphi)$ by
\begin{equation}
 \oint_\gamma A=+\,\Omega_{\mathrm{sol}}(\gamma)\qquad(\mathrm{mod}\ 2\pi) ;
 \label{eq:hol}
\end{equation}
equivalently, the coordinate frame turns by $-\Omega_{\mathrm{sol}}$ relative to the parallel one.
\end{corollary}

\begin{proof}
Integrate \eqref{eq:dpsi} along $\gamma$ and apply Stokes' theorem with $\dd A=\omega$. Two
provisos. The form $A$ is only a local representative --- the frame $(\hat e_\theta,\hat e_\varphi)$
is singular at the poles --- so if $S$ meets a pole one subdivides $S$ into pieces each contained in
a chart and applies the argument to each, the contributions of the interior edges cancelling in
pairs; what survives is the trivialisation-independent statement that the holonomy angle is
$\int_S\omega$. And $S$ itself is ambiguous: replacing it by its complement changes $\int_S\omega$
by $\int_{S^{2}}\omega=4\pi$, which does not change the rotation, whence the reduction modulo
$2\pi$.
\end{proof}

\noindent The two readings in Corollary~\ref{cor:hol} are inverse to each other, and are easily
confused. The first is the one an observer measures on the transported field, and it carries a plus
sign. Equation \eqref{eq:hol} is the rotation found by Rytov~\cite{Rytov1938} and
Vladimirskii~\cite{Vladimirskii1941}, and recognised as a geometric phase by
Berry~\cite{Berry1987,Berry1984} and Simon~\cite{Simon1983}. This section shows that it is not an
extra postulate about polarization. It is the holonomy of the connection induced by transversality,
and it follows from \eqref{eq:leak} alone.

\section{The curvature of the polarization state bundle}\label{sec:curv}

The space $Z$ of beam states has been built, its two motions named, and the rule $D$ that gives the
second of them a meaning is now available. The chart of \S5.3 gives the curvature in three terms, which are
not the vertical, mixed and horizontal parts; \S5.4 explains why and corrects it, and
Theorem~\ref{thm:split} is the corrected statement. We can therefore differentiate a phase. Lemma~\ref{lem:proj}
applies a second time, to the projector onto a single state inside $\Eperp$, with $D$ and not
$\nabla$ as the ambient connection. Since $D$ is not flat, the first term of \eqref{eq:projcurv} no
longer vanishes, and the resulting curvature carries two contributions at once.

\subsection{The Pancharatnam connection}

The connection $D$ was built on $\Eperp\to S^{2}$, while $\mathcal L$ sits inside
$\pi^{*}\Eperp\to Z$; so $D$ must first be carried to $Z$. It pulls back: there is a unique
connection
\begin{equation}
 \pi^{*}D:\ \Gamma(\pi^{*}\Eperp)\longrightarrow\Omega^{1}(Z;\pi^{*}\Eperp),\qquad
 (\pi^{*}D)_V(\pi^{*}\Phi)=\pi^{*}\big(D_{\dd\pi_\ell(V)}\Phi\big)
 \label{eq:pullD}
\end{equation}
for $\Phi\in\Gamma(\Eperp)$ and $V\in T_\ell Z$, extended to all sections of $\pi^{*}\Eperp$ by the
Leibniz rule. Vertical directions are invisible to it in the following sense: if $V$ is vertical
then $\dd\pi_\ell(V)=0$, so $(\pi^{*}D)_V(\pi^{*}\Phi)=0$ --- a field that depends only on the
propagation direction does not change when one moves inside a Poincaré sphere. A general section of
$\pi^{*}\Eperp$ does depend on the state as well, and does vary vertically.

\begin{definition}\label{def:panch}
With $\mathcal L$ as in Definition~\ref{def:taut} and
$\Pi_\ell:\Eperp(\hat k)\to\Eperp(\hat k)$ the orthogonal projector with image $\ell$, the
\emph{Pancharatnam connection} is
\begin{equation}
 \nabla^{\mathcal L}=\Pi\circ\pi^{*}D\ :\
 \Gamma(\mathcal L)\longrightarrow\Omega^{1}(Z;\mathcal L) .
 \label{eq:panch}
\end{equation}
\end{definition}

The definition is canonical once $D$ and the orthogonal projection onto $\ell$ are fixed. An amplitude in $\ell$ is transported by the ambient rule
--- which, by Section~\ref{sec:transv}, is $D$ and not $\dd$, because the plane moves --- and the
result is projected back onto the state, because $Z$ remembers only the state. This is the
``differentiate, then project'' of \S4.2 applied a second time, with $\Pi$ in the role that $P$
played there; it is the second of the two projections. On a local section, since
$\Pi P=\Pi$, \eqref{eq:panch} reads $\Pi\,\dd\psi$, the familiar Berry--Pancharatnam
connection~\cite{Berry1984,Simon1983,Pancharatnam1956}, whose holonomy around a closed circuit is
the geometric phase of the beam.

\subsection{Vertical and horizontal subspaces}

A tangent vector to $Z$ at $\ell$ is an infinitesimal change of the beam state. Two kinds are
distinguished, and they are the two physical mechanisms.

\emph{Vertical.} The direction is held fixed and the polarization moves inside its own Poincaré
sphere. These are the vectors killed by $\dd\pi$:
\begin{equation}
 V_\ell=\ker\dd\pi_\ell\ \subset\ T_\ell Z ,
 \label{eq:V}
\end{equation}
the tangent space to the fibre. This is what the medium does: birefringence and dichroism act inside
the transverse plane and do not move $\hat k$. They act on the fibre by M\"obius transformations of
the Poincaré sphere, which are rotations only in the absence of loss; nothing below depends on that
action being unitary, because the connection and its curvature are properties of $Z$, not of the
medium.

\emph{Horizontal.} The direction moves and the polarization is carried along without being turned in
its plane. This is the beam crossing a piece of the isotropic background: nothing acts on the
polarization, and yet its point moves in $Z$, because the Poincaré sphere it lives on is being
carried to another direction. The recipe is the one Section~\ref{sec:transv} already gives. Let $X\in T_{\hat k}S^{2}$ be
a direction of motion; take any curve through $\hat k$ with velocity $X$, transport $\ell$ along it
with $D$, and differentiate at $\hat k$. The result depends on $X$ alone and not on the curve, and
defines a linear map
\begin{equation}
 \mathrm{hor}_\ell:\ T_{\hat k}S^{2}\longrightarrow T_\ell Z,
 \qquad \dd\pi_\ell\circ\mathrm{hor}_\ell=\mathrm{id},
 \label{eq:hor}
\end{equation}
and we set $H_\ell=\im\,\mathrm{hor}_\ell$. Since $\mathrm{hor}_\ell$ splits $\dd\pi_\ell$,
\begin{equation}
 T_\ell Z=H_\ell\oplus V_\ell .
 \label{eq:HV}
\end{equation}
This is what bending alone does to the polarization: the Rytov transport of
Section~\ref{sec:transv}, read on states
rather than on vectors. In the language of bundles it is the standard construction: $D$ is a
principal connection on the bundle $P_U\to S^{2}$ of unitary frames of $\Eperp$, the space of
states is the associated bundle $Z=P_U\times_{U(2)}\CP^{1}$ with fibre $\CP^{1}$, and a principal
connection induces a horizontal distribution on any bundle associated with it
\cite[Vol.~I, Ch.~II, \S7]{KobayashiNomizu}. The proofs below use the horizontal lift \eqref{eq:hor} directly.
Remark~\ref{rem:minimal} places Theorem~\ref{thm:split} in this general setting.

Nothing in \eqref{eq:HV} refers to a frame or to coordinates: the splitting is made by $D$. This
matters in \S5.4, where the obvious coordinates do not respect it.

\subsection{Computation in a chart}

Fix a direction $\hat k$ away from the poles, so that the frame $(\hat e_\theta,\hat e_\varphi)$ of
Section~\ref{sec:transv} is available, and use the circular basis of the transverse plane,
\begin{equation}
 u_\pm=\tfrac1{\sqrt2}\big(\hat e_\theta\pm\ii\,\hat e_\varphi\big) .
 \label{eq:upm}
\end{equation}
Let $\Sigma\in\Gamma\big(\End(\Eperp)\big)$ be the helicity operator, acting fibrewise by
\begin{equation}
 \Sigma(\hat k):\ \Eperp(\hat k)\to\Eperp(\hat k),\qquad \Sigma\Phi=\ii\,\hat k\times\Phi .
 \label{eq:sigma}
\end{equation}
Since $\hat k\times\hat e_\theta=\hat e_\varphi$ and $\hat k\times\hat e_\varphi=-\hat e_\theta$, one
has $\Sigma u_\pm=\pm u_\pm$: in the basis $(u_+,u_-)$ the helicity is the matrix $\sigma_3$. Write
a unit amplitude as
\begin{equation}
 \psi(\vartheta,\beta)=\cos\tfrac\vartheta2\,u_+ +e^{\ii\beta}\sin\tfrac\vartheta2\,u_- .
 \label{eq:chart}
\end{equation}
Then $(\vartheta,\beta)$ are polar coordinates on the Poincaré sphere of that direction, with the
circular states at the poles and the linear states on the equator, as in the usual optical
convention; the circular Stokes parameter is the polar component,
\begin{equation}
 s_3=\langle\psi,\Sigma\psi\rangle=\cos\vartheta ,
 \label{eq:s3}
\end{equation}
a smooth function $s_3:Z\to[-1,1]$;
and on the equator $\vartheta=\pi/2$ the state \eqref{eq:chart} is, up to a phase, the linear
polarization $\cos\frac\beta2\,\hat e_\theta+\sin\frac\beta2\,\hat e_\varphi$, so that $\beta/2$ is
the azimuth of the polarization ellipse measured from $\hat e_\theta$. The four angles
$(\theta,\varphi,\vartheta,\beta)$ are a chart of $Z$, two for the direction and two for the
polarization; the state is $\ell=\CC\psi$ and its projector $\Pi=\psi\psi^{\dagger}$.

\paragraph{The area form of one fibre.} One two-form on the fibre will appear throughout, and we fix
it here. Up to a factor, a fibre $\CP^{1}=\CP\big(\Eperp(\hat k)\big)$ carries exactly one
two-form invariant under the unitary transformations of $\Eperp(\hat k)$: the K\"ahler form of the
Fubini--Study metric. In the coordinates \eqref{eq:chart} we take it to be
\begin{equation}
 \omega_{\mathrm{FS}}=\tfrac12\sin\vartheta\;\dd\vartheta\wedge\dd\beta ,
 \qquad \int_{\CP^{1}}\omega_{\mathrm{FS}}=2\pi ,
 \label{eq:FS}
\end{equation}
that is, half the area form of the unit Poincar\'e sphere. Two things fix the factor $\tfrac12$. It
is the normalisation in which the tautological bundle of a single fibre has $c_1=-1$, which is what
makes the Chern numbers of Section~\ref{sec:global} integers in the usual convention; and it is the
optical statement that a closed circuit on the Poincar\'e sphere yields \emph{half} the enclosed
area as a phase. Note that $\omega_{\mathrm{FS}}$ is a form on the fibre only: it is defined on
$V_\ell$ and extended to $T_\ell Z$ by declaring it zero whenever one of its arguments is
horizontal.

\paragraph{The connection one-form.} Equation \eqref{eq:chart} is a local unit section of
$\mathcal L$ over the chart $U_Z$, hence a local trivialisation of it. Since $\mathcal L$ has rank
one, $\nabla^{\mathcal L}\psi$ is a multiple of $\psi$, so there is a unique one-form $a$ with
\begin{equation}
 \nabla^{\mathcal L}\psi=a\,\psi ,\qquad a\in\Omega^{1}(U_Z;\ii\RR) ,
 \label{eq:adef}
\end{equation}
the connection one-form in that frame. It is computed once and for all: $\psi$ being normalised, the
orthogonal projector onto $\mathcal L_\ell=\CC\psi$ is $\Pi_\ell\Phi=\psi\langle\psi,\Phi\rangle$,
so that $\nabla^{\mathcal L}_V\psi=\Pi_\ell\big((\pi^{*}D)_V\psi\big)
=\psi\,\langle\psi,(\pi^{*}D)_V\psi\rangle$, and comparison with \eqref{eq:adef} gives
\begin{equation}
 a=\langle\psi,\pi^{*}D\,\psi\rangle .
 \label{eq:aform}
\end{equation}
That $a$ takes purely imaginary values follows from $\dd\langle\psi,\psi\rangle=0$ and the metric
compatibility of $D$. From here on we write $D$ for $\pi^{*}D$, no confusion being possible.

It remains to put \eqref{eq:aform} in coordinates, and one fact from
Section~\ref{sec:transv} suffices. In the frame $(\hat e_\theta,\hat e_\varphi)$ the induced
connection is $D=\dd-A\,J$ with $A=-\cos\theta\,\dd\varphi$, by \eqref{eq:A}; since
$Ju_\pm=\mp\ii u_\pm$, in the basis $(u_+,u_-)$ this reads
\begin{equation}
 D=\dd-\ii\cos\theta\,\dd\varphi\ \sigma_3 ,
 \label{eq:Dframe}
\end{equation}
now with $\dd$ differentiating in all four coordinates of $Z$, the pullback having changed the base
and not the fibre. Hence
\begin{equation}
 a=\psi^{\dagger}\dd\psi-\ii\cos\theta\,\dd\varphi\ \psi^{\dagger}\sigma_3\psi
 =\ii\sin^{2}\tfrac\vartheta2\;\dd\beta\;-\;\ii\,s_3\cos\theta\,\dd\varphi ,
 \label{eq:a}
\end{equation}
the first term by direct differentiation of \eqref{eq:chart}, the second by \eqref{eq:s3}.
Its exterior derivative is the curvature,
$F_{\mathcal L}=\dd a\in\Omega^{2}(Z;\ii\RR)$ --- for a line bundle the curvature is a plain
two-form, not an endomorphism-valued one --- and with
$\dd s_3=-\sin\vartheta\,\dd\vartheta$ it reads
\begin{equation}
 F_{\mathcal L}=\dd a
 =\underbrace{\tfrac{\ii}2\sin\vartheta\;\dd\vartheta\wedge\dd\beta}_{\text{polarization only}}
 \ \underbrace{-\ \ii\cos\theta\;\dd s_3\wedge\dd\varphi}_{\text{mixed}}
 \ \underbrace{+\ \ii\,s_3\,\sin\theta\;\dd\theta\wedge\dd\varphi}_{\text{direction only}} .
 \label{eq:Fchart}
\end{equation}
Three terms. Everything this section proves is already contained in them, and what remains is to
read them correctly.

\begin{figure}[t]\centering
\includegraphics[width=\textwidth]{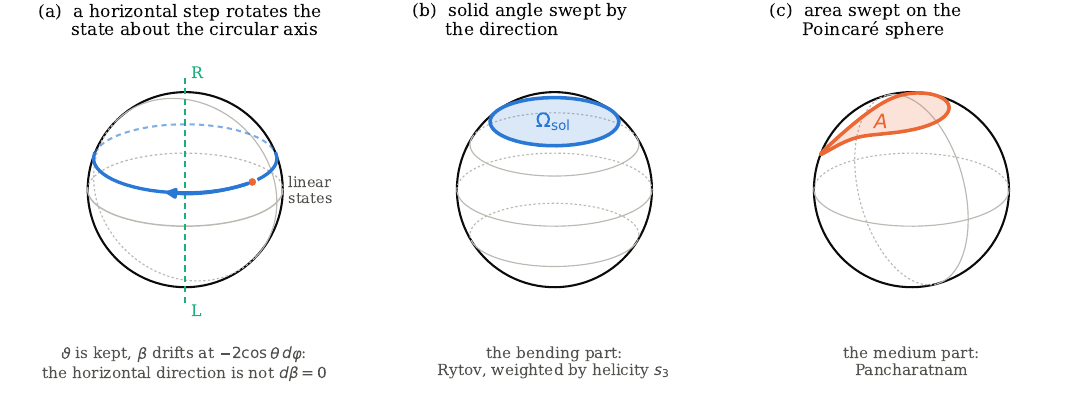}
\caption{(a) Why the chart is not adapted to the splitting. A horizontal step --- moving the
direction while carrying the state without turning it --- rotates the Poincaré sphere about its
circular axis: the latitude $\vartheta$ is kept, but the longitude $\beta$, twice the azimuth of the
ellipse measured from the moving frame $\hat e_\theta$, drifts at the rate $-2\cos\theta\,\dd\varphi$.
The horizontal directions are therefore not $\dd\beta=0$. (b), (c) What the two surviving terms of
the curvature measure: the solid angle swept by the direction, weighted pointwise by the helicity of
the state, and the area swept on the Poincaré sphere.}
\label{fig:split}
\end{figure}

\subsection{Adapted one-forms}

One is tempted to read the three terms of \eqref{eq:Fchart} as the vertical, mixed and horizontal
parts. That reading is wrong, for the reason the connection was introduced in the first place.

A horizontal step is a $D$-parallel transport, and by \eqref{eq:Dframe} it acts on the amplitude as
$\dd\psi=\ii\cos\theta\,\dd\varphi\,\sigma_3\psi$: the component on $u_+$ is multiplied by
$e^{\ii\alpha}$ and the component on $u_-$ by $e^{-\ii\alpha}$, with $\dd\alpha=\cos\theta\,\dd\varphi$.
On the Poincaré sphere this is a rotation about the circular axis (Figure~\ref{fig:split}(a)), and
in the coordinates of \eqref{eq:chart}
\begin{equation}
 \dd\vartheta\big|_{H}=0 ,\qquad \dd\beta\big|_{H}=-2\cos\theta\;\dd\varphi .
 \label{eq:hlift}
\end{equation}
This is the Rytov law read on states, and in the Stokes-vector language it is the precession about the $S_3$ axis, with
$S_3$ conserved~\cite{BliokhFrolovKravtsov2007,Bliokh2008,Bliokh2009}: the azimuth $\beta/2$ of the ellipse turns, relative to the
frame $(\hat e_\theta,\hat e_\varphi)$, by $-\cos\theta\,\dd\varphi$, which is \eqref{eq:dpsi}. At the
poles, where $\beta$ is undefined, the same step is a pure phase: bending turns a linear state on the
Poincaré sphere and only rephases a circular one.

So $\dd\beta$ does not annihilate horizontal vectors, and the first term of \eqref{eq:Fchart} is not
the vertical part. Nor does the middle term vanish on a horizontal--vertical pair $(H,V)$: since
$\dd s_3(H)=0$ by \eqref{eq:hlift} while $\dd\varphi(H)\neq0$, one has
$(\dd s_3\wedge\dd\varphi)(H,V)=-\dd s_3(V)\,\dd\varphi(H)$. Both of the first two terms contribute on
mixed pairs, and they cancel: by \eqref{eq:hlift},
$\tfrac{\ii}2\sin\vartheta\,(\dd\vartheta\wedge\dd\beta)(H,V)=\ii\sin\vartheta\cos\theta\,
\dd\vartheta(V)\,\dd\varphi(H)=-\ii\cos\theta\,\dd s_3(V)\,\dd\varphi(H)$, which is the negative of
the middle term.

The repair is to use, instead of $\dd\beta$, the one-form that \eqref{eq:hlift} annihilates,
\begin{equation}
 \delta\beta=\dd\beta+2\cos\theta\,\dd\varphi ,
 \label{eq:delta}
\end{equation}
which vanishes on $H$ by construction and reduces to $\dd\beta$ on $V$, where $\dd\varphi=0$; the
form $\dd\vartheta$ needs no repair, since it already vanishes on $H$. Expanding
$\dd\vartheta\wedge\delta\beta$ reproduces the first two terms of \eqref{eq:Fchart} exactly, so that
\begin{equation}
 \boxed{\;F_{\mathcal L}
 =\frac{\ii}2\sin\vartheta\;\dd\vartheta\wedge\delta\beta
 \;+\;\ii\,s_3\,\sin\theta\;\dd\theta\wedge\dd\varphi\;}
 \label{eq:Fsplit}
\end{equation}
The separation is now genuine, and can be checked term by term. The first term kills any pair
containing a horizontal vector, and on a pair of vertical vectors it is
$\tfrac{\ii}2\sin\vartheta\,\dd\vartheta\wedge\dd\beta=\ii\,\omega_{\mathrm{FS}}$ by
\eqref{eq:FS}. The second kills any pair containing a vertical vector, and on a pair of horizontal ones it
is $\ii s_3$ times the area form of the direction sphere. Nothing survives on a mixed pair. Two
terms, not three; the only thing the connection had to supply is the correction
$2\cos\theta\,\dd\varphi$ in \eqref{eq:delta}, twice the Rytov rate.

\subsection{Intrinsic form of the splitting}

\begin{theorem}\label{thm:split}
At each $\ell\in Z$, the decomposition \eqref{eq:HV} induces a decomposition of the two-form
$F_{\mathcal L}$. Write $F_{VV}$, $F_{HV}$, $F_{HH}$ for its restrictions to
$V_\ell\times V_\ell$, to $H_\ell\times V_\ell$ and to $H_\ell\times H_\ell$. Let
$s_3=\langle\psi,\Sigma\psi\rangle/\langle\psi,\psi\rangle$ be the circular Stokes parameter of
$\ell=\CC\psi$, with $\Sigma$ as in \eqref{eq:sigma}, and let $\omega_{\mathrm{FS}}$ be the
Fubini--Study form \eqref{eq:FS} of the fibre through $\ell$. Then
\begin{equation}
 F_{VV}=\ii\,\omega_{\mathrm{FS}} ,\qquad
 F_{HH}=\ii\,s_3\,\omega ,\qquad
 F_{HV}=0 .
 \label{eq:thmsplit}
\end{equation}
\end{theorem}

\begin{proof}
Apply Lemma~\ref{lem:proj} to the projector $\Pi$ inside the bundle $\pi^{*}\Eperp$ with ambient connection
$\pi^{*}D$:
\begin{equation}
 F_{\mathcal L}=\Pi\,(\pi^{*}F_D)\,\Pi+\Pi\,(D\Pi\wedge D\Pi)\,\Pi .
 \label{eq:nested}
\end{equation}
On two vertical arguments $\dd\pi$ vanishes, so the first term vanishes and only the second survives;
it is the curvature of the tautological bundle of a single $\CP^{1}$ with its induced Hermitian
metric, that is $\ii\,\omega_{\mathrm{FS}}$ \cite{GriffithsHarris}. On two horizontal arguments the
lift is $D$-parallel, so $D\Pi$ annihilates them and only the first term survives:
$F_{HH}(X,Y)=\langle\psi,F_D(X,Y)\psi\rangle/\langle\psi,\psi\rangle$ for $X,Y$ an oriented
orthonormal pair of $T_{\hat k}S^{2}$. By \eqref{eq:FD} and
$\psi=aX+bY$ this is $2\ii\operatorname{Im}(\bar ab)=\ii\langle\psi,\Sigma\psi\rangle$, which is
$\ii s_3$ after normalisation; multiplying by $\omega(X,Y)=1$ gives the second identity. On one
horizontal and one vertical argument both terms of \eqref{eq:nested} vanish: the first because
$\pi^{*}F_D$ needs two non-vertical arguments, the second because $D\Pi$ annihilates the horizontal
one.
\end{proof}

\begin{remark}\label{rem:minimal}
Theorem~\ref{thm:split} is the minimal-coupling form for an $S^1$-action. The curvature of $D$ takes values in
$\mathfrak{so}(2)$, generated by $J$ of \eqref{eq:A}, so that the unitary frame bundle of $\Eperp$ reduces to the
$SO(2)$-bundle of orthonormal frames of $TS^2$, and $Z=SO(3)\times_{SO(2)}\CP^1$. A rotation of the transverse
plane by the angle $\alpha$ multiplies the components on $u_\pm$ by $e^{\mp\ii\alpha}$, hence, for the active action $e^{\alpha J}$ and the state \eqref{eq:chart}, shifts $\beta$
by $2\alpha$: it acts on the Poincar\'e sphere as the flow of $2\partial_\beta$. Since
$\iota_{2\partial_\beta}\omega_{\mathrm{FS}}=\dd s_3$, this flow is Hamiltonian with moment map $s_3$. The curvature of the invariant connection on
$\mathcal L$ is then the sum of $\ii\omega_{\mathrm{FS}}$ and of the curvature of $D$ paired with this moment
map, which is $\ii s_3\omega$, and the mixed part vanishes because the horizontal distribution is that of the
principal connection \cite{Sternberg1977,Weinstein1978}. The direct proof above establishes the splitting for this bundle with its explicit ingredients: the identification of the connection with the
complexified Levi-Civita connection, and of the moment map with the helicity. The vanishing of the mixed part follows from the
parallelism of $\Pi$ along horizontal vectors.
\end{remark}

\noindent The proof is short because Lemma~\ref{lem:proj} was stated for a general projector field
and a general ambient connection. This is in what sense transversality and the passage to states are
one construction and not two: the same lemma is applied twice, with $\nabla$ flat the first time and
$D$ curved the second, and the curvature left over from the first projection is what produces the
term $F_{HH}$ of the second.

\subsection{Geometric phase of a beam}

\paragraph{Holonomy and geometric phase.} The curvature is not itself a phase: $F_{\mathcal L}$ takes
values in $\ii\RR$. With the convention \eqref{eq:a}, a section $f\psi$ of $\mathcal L$ is parallel
when $\dd f+af=0$, so the holonomy around a closed circuit $\gamma$ of beam states --- a beam whose
direction and polarization state both return to their initial values --- is
$\exp(-\oint_\gamma a)$. If $\gamma$ bounds a surface $S\subset Z$, Stokes' theorem gives
\begin{equation}
 \mathrm{Hol}_\gamma=\exp\Big(-\int_S F_{\mathcal L}\Big)=e^{\ii\gamma_{\rm g}},\qquad
 \gamma_{\rm g}=\ii\int_S F_{\mathcal L}\quad(\mathrm{mod}\ 2\pi),
 \label{eq:hol3}
\end{equation}
and by Theorem~\ref{thm:split}
\begin{equation}
 \gamma_{\rm g}=-\int_S\omega_{\mathrm{FS}}\;-\;\int_S s_3\,\pi^{*}\omega\qquad(\mathrm{mod}\ 2\pi).
 \label{eq:phase}
\end{equation}
The first integral is the area swept on the Poincaré sphere, the second the solid angle swept by the
propagation direction, weighted at each point by the helicity of the state carried there; for a circular
state it is the Berry phase of the circular modes in the isotropic case~\cite{BliokhFrolovKravtsov2007,Bliokh2008,Bliokh2009}
(Figure~\ref{fig:split}(b,c)). Only their sum is intrinsic. Two surfaces with the same boundary differ by a closed surface, on
which the total integral is a multiple of $2\pi$ by the integrality of $c_1$; the two terms
separately need not be, since neither is a closed form by itself. The splitting of
Theorem~\ref{thm:split} is a statement about the curvature at each point, not about how an
integrated phase is shared between the two mechanisms along a given surface.

\paragraph{Geometric phase and the physical field.} The connection $\nabla^{\mathcal L}$ is Hermitian, so its holonomy
is unitary, also when $\chi$ is not Hermitian. The medium enters through the curve of states along which the holonomy is
computed, and not through the connection. The physical field carries in addition a phase and a variation of norm
generated by the material term of \eqref{eq:fulltransport}, which are not part of $\gamma_{\rm g}$. For $\chi=\lambda(\bm x)\Id$ the curve in $Z$ is the one obtained for
$\chi=0$, whereas the amplitude $\bm e$ is multiplied by $\exp\big(\tfrac{\ii}{2}\int\lambda/n_0\,\dd\sigma\big)$, which is complex
when $\lambda$ is.

\begin{figure}[t]\centering
\includegraphics[width=0.84\textwidth]{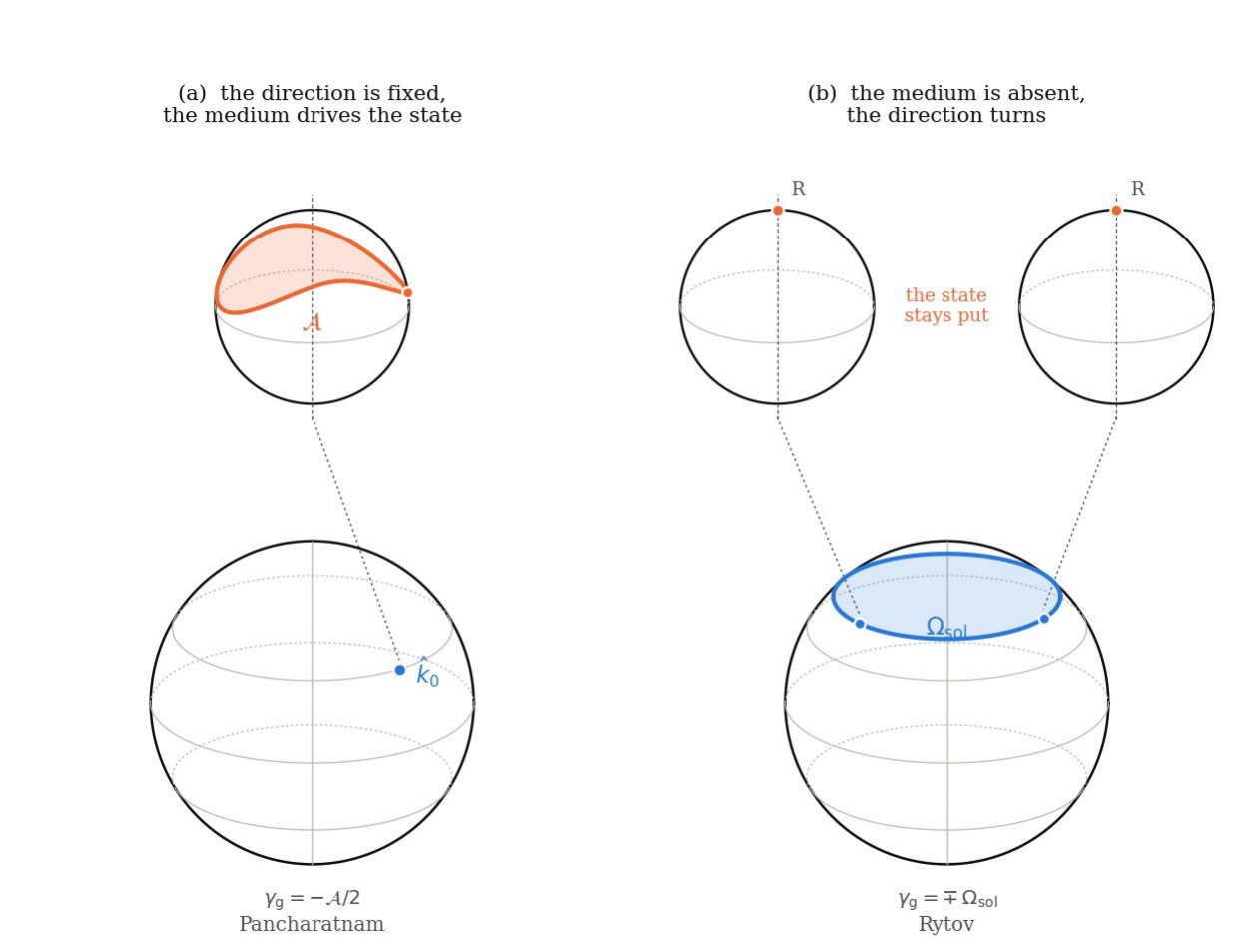}
\caption{The two circuits for which one of the two terms of \eqref{eq:phase} vanishes. (a) The
direction stays at $\hat k_0$ and the medium takes the polarization state around a loop on the
Poincaré sphere of that direction, enclosing the area $\mathcal A$. (b) The medium is absent and
the direction goes around a loop enclosing the solid angle $\Omega_{\mathrm{sol}}$, while the state
stays circular, that is, at the same pole of every Poincaré sphere along the way. In (a) the phase
is $-\mathcal A/2$, in (b) it is $\mp\Omega_{\mathrm{sol}}$.}
\label{fig:cases}
\end{figure}

\paragraph{Two special cases.} Equation \eqref{eq:phase} has two terms. There are two experiments
in which one of the two vanishes, and each gives a phase that was known before. Figure~\ref{fig:cases}
shows the two circuits.

\emph{First case: the direction is fixed and the medium acts.} The beam propagates in the same
direction $\hat k_0$ from beginning to end. Only its polarization state moves, and it comes back to
its initial value. The circuit lies in the single fibre over $\hat k_0$, so we may take $S$ inside
that fibre too. On such an $S$ the second term vanishes: $\pi$ sends $S$ to the one point $\hat k_0$,
so $\pi^{*}\omega$ is zero on $S$. The first term is the area of $S$ measured by
$\omega_{\mathrm{FS}}$, which by \eqref{eq:FS} is half the area $\mathcal A$ enclosed by the circuit
on the unit Poincaré sphere. Hence
\begin{equation}
 \gamma_{\rm g}=-\tfrac12\mathcal A .
 \label{eq:pancharatnamphase}
\end{equation}
This is the phase of Pancharatnam~\cite{Pancharatnam1956,Berry1987b}.

\emph{Second case: the medium is absent and the direction turns.} Now the state is transported
horizontally. One point has to be settled first. By \eqref{eq:hlift}, for a generic loop of directions a horizontally transported state
returns to its starting point in $Z$ only if it is circular; a linear state comes back with its
azimuth turned by the Rytov angle, so its circuit is not closed and \eqref{eq:hol3} does not apply to
it. If $\Omega_{\mathrm{sol}}\in\pi\ZZ$, the longitude shifts by $2\Omega_{\mathrm{sol}}\equiv0\pmod{2\pi}$ and
every state closes: a cycle on the sphere of directions corresponds to a rotation by $2\Omega_{\mathrm{sol}}$ of the Poincar\'e sphere~\cite{Bliokh2009}. Take the state circular, that is, equal to $\ell_\pm(\hat k)$ at each direction. The section
$\ell_\pm$ is $D$-parallel, so its image over a region $S_{\rm dir}$ of the direction sphere is a
horizontal surface; take $S$ to be that image. The first term vanishes on it, because
$\omega_{\mathrm{FS}}$ is a form on the fibre, by \eqref{eq:FS}, and $S$ has no vertical
directions. In the second term
$s_3=\pm1$ is constant, and what is left is the solid angle:
\begin{equation}
 \gamma_{\rm g}=\mp\,\Omega_{\mathrm{sol}} .
 \label{eq:rytovphase}
\end{equation}
This is the phase of a circularly polarized beam whose direction is carried around a closed loop. It
was predicted by Chiao and Wu and measured by Tomita and Chiao in a helically wound optical
fibre~\cite{Chiao1986,Tomita1986}, and it is the phase counterpart of the rotation of
Corollary~\ref{cor:hol}: there a linear state turns by $\Omega_{\mathrm{sol}}$, here a circular state
acquires the phase $\mp\Omega_{\mathrm{sol}}$, which is the same statement read on the two kinds of
state.

\emph{In general} both terms of \eqref{eq:phase} contribute, each depending on the surface chosen. If the state is neither circular nor at a fixed
direction, the circuit does not close in general, and the phase is the non-cyclic
one of Samuel and Bhandari~\cite{SamuelBhandari1988}. It is obtained by closing the curve with a geodesic
of the fibre, which requires the initial and final directions to coincide; the phase is defined only if the
initial and final states are not orthogonal, since the Pancharatnam overlap vanishes otherwise.

\paragraph{Vanishing of the mixed curvature.} The two phases are usually presented as two separate mechanisms
that happen to coexist. The vanishing of the mixed curvature says more. Take a path $c$ of directions of length
$\varepsilon$ (bending), and a vertical vector field $Y$ on $Z$, the generator of the medium. The action of the medium is the flow of $Y$
for a duration $\delta$, the same in the two steps where it acts. The circuit
``bend, act, unbend, undo'' is built as follows. Lift $c$ horizontally from $\ell_0$ to $\ell_1$; follow the action of
the medium from $\ell_1$ to $\ell_2$ in the fibre over $c(\varepsilon)$; lift $c$ backwards horizontally from $\ell_2$ to
$\ell_3$ in the fibre over $c(0)$; follow the inverse action of the medium over $c(0)$ from $\ell_3$ to a point $\ell_4$.
This does not close in general, because the medium acts at two different directions and its action is not transported
by $D$ into itself. The point $\ell_4$ differs from $\ell_0$ by $O(\varepsilon\delta)$ within the fibre, and we close the
circuit by the geodesic of the fibre from $\ell_4$ to $\ell_0$.

\begin{proposition}\label{prop:commute}
Let $\tau$ denote the $D$-parallel transport of the fibre over $c(\varepsilon)$ to the fibre over $c(0)$ along $c$, and assume
\begin{equation}
 \|\tau_{*}Y_{c(\varepsilon)}-Y_{c(0)}\|\le C\,\varepsilon .
 \label{eq:Hyp}
\end{equation}
Here $\varepsilon$ is the length of $c$. The inequality \eqref{eq:Hyp} is a hypothesis on the protocol of the medium: it compares, in one fibre,
the generators acting at the two ends of $c$. It holds with $C$ depending on bounds on $Y$, its derivatives and the curvature when $Y$ is a
smooth vertical vector field on $Z$. In the Maxwell model the generator is built from the transverse block of $\chi$, the factor $1/n_0$ and the
transport $D$ used to compare the two planes, so that \eqref{eq:Hyp} involves all three and not the variation of $\chi$ alone; two positions
with the same direction may carry different generators. Then the holonomy of $\nabla^{\mathcal L}$ around the closed circuit $\gamma_{\varepsilon,\delta}$ so defined
satisfies $\mathrm{Hol}(\gamma_{\varepsilon,\delta})=1+O(\varepsilon\delta^2)$, with a constant depending on $C$ and on the same bounds.
\end{proposition}

\begin{proof}
Let $R$ be the surface swept by transporting horizontally the curve $\ell_1\ell_2$ along $c$ back to $c(0)$. Horizontal
lifts map fibres to fibres, so every transported curve is vertical, and $T R$ is spanned at each point by a horizontal and a
vertical vector. By Theorem~\ref{thm:split}, $F_{\mathcal L}$ vanishes on $TR$, and the holonomy around $\partial R$ is $1$.
The boundary $\partial R$ consists of the first two steps, the third step, and the curve $\kappa$ in the fibre over $c(0)$ from $\ell_0$
to $\ell_3$ obtained by transporting $\ell_1\ell_2$. The circuit $\gamma_{\varepsilon,\delta}$ differs from $\partial R$ in its last
side: instead of $\kappa^{-1}$ it follows the inverse action of the medium over $c(0)$ and the closing geodesic. The two
curves in the fibre over $c(0)$ are integral curves, of length $O(\delta)$, of vector fields that differ by $O(\varepsilon)$ by \eqref{eq:Hyp}, namely the
generator of the medium over $c(0)$ and the generator over $c(\varepsilon)$ transported to $c(0)$. Both curves
end at $\ell_3$. Parametrize them from $\ell_3$ by the time $t\in[0,\delta]$ of the backward flow; they reach $\ell_0$ and $\ell_4$ at
$t=\delta$. By \eqref{eq:Hyp} and Gronwall's lemma, with uniform bounds on the fields and their derivatives, their distance at time $t$ is
$O(\varepsilon t)$. The ribbon formed by the short fibre geodesics joining the points of equal time covers the region between the two
curves, and its last geodesic is the closing segment from $\ell_4$ to $\ell_0$. Its $\omega_{\mathrm{FS}}$-area is bounded by
$\int_0^\delta O(\varepsilon t)\,\dd t=O(\varepsilon\delta^2)$, the geodesics having length $O(\varepsilon t)$ and moving with speed $O(1)$. By Theorem~\ref{thm:split}, $F_{VV}=\ii\omega_{\mathrm{FS}}$ on the fibre, so this
region contributes a phase $O(\varepsilon\delta^2)$, and the claim follows.
\end{proof}

\noindent For $\varepsilon,\delta=O(\rho)$ the phase is $O(\rho^3)$, whereas circuits made of two bending steps or of two medium
steps give phases of order $\rho^2$. The closing step is part of the statement: without it the circuit is not a loop in $Z$, and
the holonomy is not defined. This is a property of the connection $D$ induced by transversality. It is invisible in the obvious
coordinates: as \S5.4 showed, two of the three terms of \eqref{eq:Fchart} are then non-zero on mixed
pairs and cancel.

\section{Chern numbers of the helicity line bundles}\label{sec:global}

The two helicity eigenstates --- the poles $\vartheta=0,\pi$ of the chart \eqref{eq:chart} ---
define sections $\hat k\mapsto\ell_\pm(\hat k)=\CC u_\pm$ of $Z$ away from the poles of the direction
sphere,
and these extend to global sections of $Z$ because the ambiguity is a phase, which the projective
fibre discards. Since $\Sigma u_\pm=\pm u_\pm$, $s_3=\pm1$ on $\ell_\pm$; this, with
\eqref{eq:sigma}, is what fixes the signs below. Let $L_\pm=\ell_\pm^{*}\mathcal L$ be the pullbacks of the tautological bundle,
line bundles over $S^{2}$.

The Chern numbers of these helicity line bundles are
known~\cite{PalmerducaQin2024,PalmerducaQin2025}. We obtain them here as the integrals of the restrictions of the
curvature of Theorem~\ref{thm:split} to the two helicity sections.

\begin{remark}\label{rem:TS2}
The integers can also be read from the identification $\Eperp=TS^2\otimes\CC$. The transverse plane at
$\hat k$ is $T_{\hat k}S^2$, the rotation by $+\pi/2$ in the plane is $\hat k\times$, and
$u_-\propto\hat e_\theta-\ii\hat e_\varphi$ spans $T^{1,0}S^2$ while $u_+$ spans $T^{0,1}S^2$. Hence
$c_1(L_-)=c_1(TS^2)=\chi(S^2)=2$ and $c_1(L_+)=-2$, in agreement with Theorem~\ref{thm:chern}. This is also the comparison with the
bundle over the light cone of~\cite{PalmerducaQin2024,PalmerducaQin2025}: the two bundles agree on each ray
through the origin, so that the radial projection identifies them. Palmerduca and Qin integrate the Berry curvature $\pm\ii\omega$ of the helicity
lines~\cite[Theorem~18]{PalmerducaQin2024}. In Theorem~\ref{thm:chern} these two curvatures are the restrictions to the two helicity
sections of the single curvature $F_{\mathcal L}$ on $Z$.
\end{remark}

\begin{figure}[t]\centering
\includegraphics[width=0.95\textwidth]{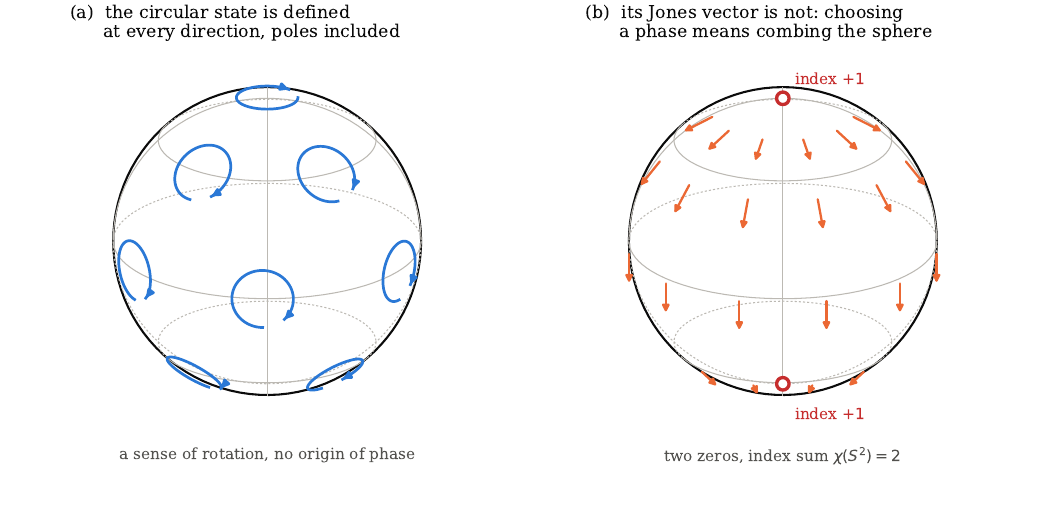}
\caption{What the two integers forbid. (a) At every propagation direction, the right-circular
\emph{state} is defined: it is a sense of rotation in the transverse plane, and the poles of the
direction sphere are no exception. (b) A Jones \emph{vector} for that state is a choice of phase
origin, that is, of a unit vector in the transverse plane: a comb of the direction sphere. One comb is drawn, with two zeros of index $+1$;
for any comb the indices of the zeros sum to $\chi(S^{2})=2$. The state is global, its phase is not, and the
obstruction is the integer that Theorem~\ref{thm:chern} computes.}
\label{fig:comb}
\end{figure}

First the statement that the two integers make precise. Pick the right-circular state at every
direction. This is legitimate, because ``right-circular'' names a sense of rotation in the
transverse plane and needs no reference axis (Figure~\ref{fig:comb}(a)). Now try to attach to it an
actual Jones vector, that is, an amplitude with a definite phase. Fixing the phase of $u_+$ is the same as
choosing a unit vector in the transverse plane, since by \eqref{eq:upm} the real part of
$e^{\ii\alpha}u_+$ is $(\cos\alpha\,\hat e_\theta-\sin\alpha\,\hat e_\varphi)/\sqrt2$; doing it
smoothly at every direction is combing the sphere of directions, which cannot be done
(Figure~\ref{fig:comb}(b)). The rest of this section turns that impossibility into a number.

\begin{theorem}\label{thm:chern}
With the orientation and the sign convention \eqref{eq:sigma} fixed above,
\begin{equation}
 c_1(L_+)=-2,\qquad c_1(L_-)=+2,\qquad c_1(L_+)+c_1(L_-)=0 ,\qquad
 c_1\Hom(L_-,L_+)=-4 .
 \label{eq:chern}
\end{equation}
\end{theorem}

\begin{proof}
On $\ell_\pm$ one has $s_3=\pm1$, so Theorem~\ref{thm:split} gives
$F_{\mathcal L}|_{\ell_\pm}=\pm\ii\,\omega$ and
\begin{equation*}
 c_1(L_\pm)=\frac{\ii}{2\pi}\int_{S^{2}}F_{\mathcal L}\Big|_{\ell_\pm}
 =\frac{\ii}{2\pi}\int_{S^{2}}(\pm\ii)\,\omega=\mp\frac{4\pi}{2\pi}=\mp2 ,
\end{equation*}
the same normalisation of $c_1$ that gives $-1$ for the tautological bundle of a single fibre,
fixed in Theorem~\ref{thm:split}. The remaining statements follow since $c_1$ is additive on direct
sums and $c_1\Hom(L,L')=c_1(L')-c_1(L)$ \cite[\S14]{MilnorStasheff}.
\end{proof}

\begin{corollary}\label{cor:global}
$\Eperp=L_+\oplus L_-$ is smoothly trivial, and admits a global complex frame; but neither $L_+$ nor
$L_-$ admits a nowhere-vanishing global section, so no global frame is everywhere the helicity
eigenframe. The right- and left-handed \emph{states}, being the global sections $\ell_\pm$ of $Z$,
are globally defined; the corresponding \emph{Jones vectors} are not. There is no global smooth
azimuth.
\end{corollary}

\begin{proof}
$c_1(\Eperp)=c_1(L_+)+c_1(L_-)=0$, and a rank-two complex bundle over $S^{2}$ is classified by its
clutching function up to homotopy, that is by $\pi_1(U(2))=\ZZ$, detected by the winding of the
determinant \cite[Sect.~18]{Steenrod}; here the clutching function is $\mathrm{diag}(z^{2},z^{-2})$, of
unit determinant and null winding, so $\Eperp$ is trivial. A line bundle over $S^{2}$ has a
nowhere-vanishing section if and only if $c_1=0$, which excludes $L_\pm$. There is no contradiction
with the existence of the sub-bundles $L_\pm\subset\Eperp$: the inclusion of a sub-bundle is a section of
$\Hom(L_\pm,\Eperp)$, not of $L_\pm$. The last statement is
$e(TS^{2})=\chi(S^{2})=2\neq0$, the Poincaré--Hopf obstruction to a nowhere-vanishing tangent field,
equivalently $c_1\Hom(L_-,L_+)=-4\neq0$. A global azimuth would give a global section of the unit circle bundle of $\Hom(L_-,L_+)$, namely
the relative phase of the $u_+$ and $u_-$ components, whose angle is twice the azimuth; such a section does not exist since
$c_1\Hom(L_-,L_+)\neq0$. This is Figure~\ref{fig:comb}(b).
\end{proof}

\begin{remark}
Corollary~\ref{cor:global} is the correction of a statement that is easy to make loosely. That
$\Eperp$ is trivial does \emph{not} license a global circular Jones frame; it licenses a global
complex frame, which cannot be adapted to helicity everywhere. The circular basis is nonetheless
used globally in the literature, and legitimately so, in the projective sense of the corollary.
\end{remark}

\section{Conclusion}

The content of this work can be stated in a few lines. The transverse plane of a bending ray is not
a frame we choose to work in: it is what Maxwell leaves available. At leading order the high-frequency ansatz forces
$\nabla S\cdot\bm a_0=0$ in \eqref{eq:eikonal}, so at a given propagation direction the amplitudes
that solve the equation at all are exactly those in the plane orthogonal to $\hat k$ --- the kernel,
on the characteristic set, of the principal symbol of Maxwell. The bundle $\Eperp$ is that constraint
read over the whole sphere of directions. Its derivative induced by the ambient one, projected, is the
connection for which the transported field does not rotate about the ray (Proposition~\ref{prop:unique}), and the
solvability condition of Maxwell's equations shows that this is the transport law. That derivative is
Rytov's law, Proposition~\ref{prop:curv} and
Corollary~\ref{cor:hol}, and the part it discards --- the second fundamental form
\eqref{eq:IIexplicit} --- is the longitudinal component that turning a ray forces on the derivative
of a transverse field. Projecting a second time, now onto a single polarization state and with the
curved $D$ in the ambient role, produces one two-form on the space of beam states. The horizontal
distribution defined by the first projection cuts that two-form in two, Theorem~\ref{thm:split},
with nothing left in between. Bending and the medium each contribute, and the circuit that bends, acts, unbends and undoes, closed in the fibre, has a
phase $O(\varepsilon\delta^2)$.

The known helicity Chern numbers express a topological obstruction: a circular state can
be named at every propagation direction, but a smooth circular Jones vector cannot be
chosen globally. In the present construction this follows from the restriction of the
curvature to the helicity sections, Theorem~\ref{thm:chern} and Corollary~\ref{cor:global}.

\paragraph{Scope and continuation.} The construction is carried out over $S^2$, the base over which the projected connection is defined: on the characteristic
set $\mathrm{Char}$, with $q(\bm x,\bm\xi)=\bm\xi/|\bm\xi|$, the kernel bundle of the symbol is $q^{*}\Eperp$ and its projected connection is
$q^{*}D$. The transport law \eqref{eq:fulltransport} also contains the generator $\ii P\chi(\bm x)P/(2n_0(\bm x))$, which depends on the
position along the ray, as does the ray itself through \eqref{eq:fermat}. Once the ray is given, the pair $(\hat k(s),\ell(s))$ is a
curve in $Z$, and its geometric phase is computed in $Z$ from the curvature of Theorem~\ref{thm:split}, whatever the dependence of $\chi$ on
position. Position- and wavelength-dependent corrections to the transport law, such as the transverse shift of the ray
associated with the helicity~\cite{Onoda2004}, occur at higher order in $\eta$. Extending the sub-bundle construction to that order, over a
bundle on the characteristic set rather than over $S^{2}$, is a natural continuation, in which one asks whether the correction is again a second
fundamental form.
The connection $D$ and the curvature of Theorem~\ref{thm:split} are independent of $\chi$; the medium enters through the curve of states. With
absorption, the holonomy of $\nabla^{\mathcal L}$ along this curve is unitary, and the attenuation and the material phase of the physical
field are carried by the transport law (see \S5.6).

\end{document}